%% file: main.tex
\documentclass[manuscript]{acmart}

\AtBeginDocument{%
  }

\setcopyright{acmcopyright}
\copyrightyear{2023}
\acmYear{2023}
\acmDOI{XXXXXXX.XXXXXXX}

\acmConference[ACM Comput. Surv.]{Make sure to enter the correct
  conference title from your rights confirmation emai}{Nov 11--08,
  2023}{NY}
\acmPrice{15.00}
\acmISBN{978-1-4503-XXXX-X/18/06}

\usepackage[most]{tcolorbox}
\usepackage{array}
\usepackage{caption}
\usepackage{subcaption}
\usepackage{blindtext}
\usepackage{color,soul}

\definecolor{keygenclr}{rgb}{0.95, 0.52, 0.0}
\definecolor{cmnonceclr}{rgb}{0.06, 0.75, 0.99}
\definecolor{cmpchlclr}{rgb}{0.2, 0.8, 0.2}
\definecolor{siggenclr}{rgb}{1.0, 0.0, 1.0}
\definecolor{sigvrfclr}{rgb}{0.84, 0.23, 0.24}

\usepackage{MnSymbol}

\usepackage{fontawesome5}
\usepackage{amsmath}
\usepackage{amsthm}
\usepackage{mathtools}
\usepackage{footnote}
\usepackage{graphicx}

\makeatletter
\newcommand{\inlineitem}[1][]{%
  \ifnum\value{enumi}>0
    \ifnum\value{enumii}=0
      \refstepcounter{enumi}%
    \else
      \refstepcounter{enumii}%
    \fi
    \textbf{\theenumi.} #1\hspace{\labelsep}%
  \fi
}
\makeatother

\theoremstyle{definition}
\newtheorem{definition}{Definition}[subsection]
\usepackage{multirow}
\usepackage{multicol}
\usepackage[absolute,overlay]{textpos}
\usepackage{xcolor}
\usepackage{tcolorbox}
\usepackage[perpage]{footmisc}

\usepackage{kcryptex}
\usepackage{kcryptdefs}
\usepackage{speccmds}

\usepackage{algorithm}
\usepackage{algpseudocode}
\usepackage{pdfpages}

\algtext*{EndWhile}
\algdef{SE}[DOWHILE]{Do}{doWhile}{\algorithmicdo}[1]{\algorithmicwhile\ #1}%
\usepackage{pifont}
\newcommand{\cmark}{\ding{51}}%
\newcommand{\xmark}{\ding{55}}%
\usepackage{threeparttable}

\usepackage[inline]{enumitem}

\begin{document}

\title{A Comprehensive Survey of Threshold Digital Signatures: NIST Standards, Post-Quantum Cryptography, Exotic Techniques, and Real-World Applications}

\author{Kiarash Sedghighadikolaei}
\email{kiarashs@usf.edu}
\author{Attila Altay Yavuz}
\email{attilaayavuz@usf.edu}
\affiliation{%
  \institution{University of South Florida}
  \streetaddress{4202 E. Fowler Avenue}
  \city{Tampa}
  \state{FL}
  \country{USA}
  \postcode{33620}
}

\begin{abstract}

\input{bodies/abstract}

\end{abstract}




\received{15 June 2023}
\received[revised]{15 June 2023}
\received[accepted]{15 June 2023}

\maketitle

\section{Introduction}
\input{bodies/introduction}

\input{bodies/other_works}

\subsection{Our Contribution}
\input{bodies/contribution}

\vspace{1.5mm}
\textbf{The Structure of our Survey}: Section \ref{sec:preliminaries} gives preliminaries such as the mathematical background, definitions, and security models of our target cryptographic primitives. Section \ref{sec:schnorr} examines the thresholding of Schnorr-like digital signatures including conventional-secure standardized schemes (e.g., ECDSA, EdDSA) as a major line of threshold EC/DLP-based (See \ref{def:dlp} and \ref{def:ecdlp}) signature family. In Section \ref{sec:pairing}, the construction of threshold signature schemes using cryptographic pairings, specifically the threshold BLS signature~\cite{blsboneh2004short}, is discussed. Section \ref{sec:rsa} analyzes the thresholding of RSA as the major (standardized) line of threshold schemes based on the Integer Factorization problem (See \ref{def:IntFactP}). Section \ref{sec:genericmpc} explores the thresholding of digital signature schemes using Multi-party Computation (MPC) techniques~\cite{brandao2023nist}. Section \ref{sec:PQC} investigates the thresholding of post-quantum resistant schemes, including NIST's standardized signature schemes and non-standardized proposals. Section \ref{sec:specialschemes} covers exotic signatures (e.g., Group \cite{boneh2022threshold}, Ring \cite{avitabile2023extendable}, Multi~\cite{nick2021musig2}) with their properties, threshold and PQ-secure constructions. Finally, Section \ref{sec:applications} provides a comprehensive exploration of the practical applications threshold signatures with potential take-aways.

\section{Preliminaries} \label{sec:preliminaries}
\input{bodies/notations}
\input{bodies/digital_signatures}
\input{bodies/threshold_digital_signatures}

\input{bodies/MPC}


\input{bodies/conventional_approaches}

\section{Generic Thresholding Approaches} \label{sec:genericmpc}
\input{bodies/generic_approaches}

\section{Post-quantum Secure Threshold Digital Signatures} \label{sec:PQC}
\input{bodies/PQC}

\section{Exotic (Special) Distributed Digital Schemes} \label{sec:specialschemes}
\input{bodies/special_schemes}

\section{Potential Applications and Directions for Threshold Signatures} \label{sec:applications}
\input{bodies/applications}

\section{Conclusion}
\input{bodies/conclusion}

\begin{acks}
This research is supported by the unrestricted gift from the Cisco Research Award (220159), and the NSF CAREER
Award CNS-1917627.
%
\end{acks}

\bibliographystyle{ACM-Reference-Format}
\bibliography{sample-base}

\appendix

\end{document}

%% file: bodies/abstract.tex
Threshold digital signatures enable a distributed execution of signature functionalities and will play a crucial role in the security of emerging decentralized next-generation networked systems and applications. In this paper, we provide a comprehensive and systematic survey of threshold and distributed signatures with advanced features. Our survey encompasses threshold signatures in conventional and post-quantum cryptography (PQC) settings and captures custom-design and standard signatures (e.g., conventional NIST and NIST-PQC). We examine both generic (via secure multi-party computation) and custom thresholding techniques for a myriad of signature families while investigating exotic signatures, real-life applications, and potential future research direction.

%% file: bodies/introduction.tex
In the modern digital era, ensuring the integrity, authenticity, and non-repudiation of digital information is of utmost importance. Digital signatures play a critical role in achieving these goals via using the Public Key Infrastructure (PKI) \cite{joint2020security}. To establish standardization, the {National Institute of Standards and Technology (NIST)} \cite{chen2023digital} has provided guidance on a suite of algorithms for developing standardized digital signatures. These standardized schemes include but are not limited to, ECDSA by ANSI \cite{ECDSAANSI}, EdDSA \cite{bernstein2012high,josefsson2017edwards}, RSA \cite{jonsson2003public,moriarty2016pkcs}, and pairing-based signatures. Notable efforts made to enable efficient implementation of Discrete Logarithm Problem (DLP)-based schemes such as Diffie-Hellman and Schnorr-like signatures\cite{schnorr1990efficient} on  Elliptic Curves (EC) (e.g.,  \cite{costello2016schnorrq,bernstein2012high,chen2023recommendations}), which offer efficient deployments in domains such as the automotive supply chain, Internet of Things (IoT)~\cite{yavuz2013eta,OSLO:2023}, wireless networks~\cite{7417611} and others. These digital signatures offer distinct advantages and characteristics, allowing users to choose the most suitable option based on their security and application needs. Despite their merits, these efficient digital signatures and standards are at a high risk of not fulfilling their goals in the emerging next-generation (NextG) distributed networked systems, especially in the wake of post-quantum-era~\cite{NISTPQC:Test:Automative:2021,SATIN:9193893,Yavuz:HQKDML:TPS:2022}.

The advent of quantum computers poses a significant risk to various sectors, including finance, IoT, government, and defense. The concern stems from the vulnerability of cryptographic systems, specifically digital signatures, which rely on hard computational problems that quantum computers can efficiently solve (e.g., EC/DLP). Shor's algorithm \cite{shor1999polynomial} is a quantum algorithm that utilizes the parallelism and superposition properties of quantum computers to solve these problems much faster than classical computers, thereby endangering the security of the mentioned domains. The threat becomes more evident considering the significant rise in private investments in the quantum computing industry as the International Data Corporation (IDC) forecasts that this market will continue to grow and estimates it to reach a value of \$8.6 billion by 2027. Moreover, prominent entities such as Google's Bristlecone, China's Zuchongzi and Jiuzhang 2.0, and IBM's Osprey are actively pushing the boundaries of quantum computing. In response to the aforementioned threat, the field of Post-Quantum Cryptography (PQC) has emerged, aiming to develop novel cryptographic algorithms that can withstand attacks from both classical and quantum computers. As a result, many organizations started a standardization process for post-quantum resistant cryptographic algorithms where NIST took the lead by initiating a standardization process in 2017 \cite{nistCallProposals}, resulting in the selection of four schemes in 2022 \cite{nistSelectedAlgorithms}: CRYSTALS-Kyber \cite{avanzi2017crystals} for public-key encryption and key establishment, CRYSTALS-Dilithium \cite{ducas2018crystals}, FALCON \cite{fouque2019fast}, and SPHINCS+ \cite{aumasson2019sphincs} for digital signatures. PQC is currently a highly active research area, wherein various new PQC signatures are built with extended features (Group \cite{chaum1991group}, Ring \cite{rivest2001leak} signatures, and Multi-signatures \cite{itakura1983public}), and NIST initiated a new standardization effort for additional PQC digital signatures \cite{nistStandardizationAdditional}.

Another critical limitation of standard digital signatures, whether PQ-safe or not, is that they operate within a centralized setting. That is, they rely on a single entity to manage cryptographic secrets and operations. However, this centralized approach poses risks as one party's compromise or malicious behavior can threaten the system's security. To enhance security, resilience, and robustness, \textit{Threshold cryptography} distributes the secrets and computations (e.g., using Shamir's secret sharing \cite{shamir1979share}, Multi-party computation (MPC) \cite{damgaard2012multiparty}) such that cryptographic operations are required to be done collaboratively by a sufficient number of parties. The recognized benefits have made NIST emphasize using multi-party threshold cryptographic systems using Distributed PKIs (DPKI) and announced a call~\cite{brandao2023nist} for submissions of multi-party threshold schemes. This helps NIST shape its future recommendations and guidelines by analyzing the submitted schemes and finding the best approaches, best practices, and reusable blocks for further incorporation in various applications such as Cryptocurrency \cite{gennaro2016threshold}, Blockchain \cite{pan2022multi}, IoT \cite{kurt2021lngate}, etc. This call targets schemes grouped into two categories: the first category consists of the NIST-specified primitives in various cryptographic domains, and the second category consists of non-NIST-selected schemes resembling those in the first category plus advanced functionalities. It is worth highlighting that in light of the significance and standardization of post-quantum resistant schemes, the development of threshold variants for these schemes is equally essential. Notably, Cozzo et al. \cite{cozzo2019sharing} have conducted comprehensive research in this direction, presenting a comprehensive analysis of thresholding the candidates from NIST's second round of standardization and explaining the general possible approaches along with their complexities and costs.

In addition to threshold digital signatures, other models of distributed signatures with special features, also called {\em exotic signatures}, such as Group signatures \cite{chaum1991group}, Ring signatures \cite{rivest2001leak}, and Multi-signatures \cite{itakura1983public} exist. These exotic signatures offer distributed security under different settings that are applicable in various domains such as Blockchain \cite{boneh2018compact}, Cryptocurrency \cite{kara2023efficient}, IoT \cite{li2018creditcoin}, Cloud \cite{wang2020flexible}, and others. 

In conclusion, both threshold cryptography and the aforementioned exotic signatures play a vital role in real-world applications by improving security, resilience, and robustness through privacy-preserving collaborative computation. In the subsequent section, we present an overview of existing state-of-the-art surveys in the literature that specifically examine the signatures discussed thus far.

%% file: bodies/other_works.tex
\begin{table}[H]
\tiny
  \caption{Summary of Published Surveys on Threshold and Exotic Signatures (Conventional and PQ-secure)}
  \label{tab:publishedSurveys}
  \begin{tabular}{p{0.8cm} p{0.5cm} p{4.5cm} p{2cm} p{2.5cm} p{2.5cm}}
    \toprule
    Work by & MPC \footnote{1} & \begin{tabular}{ccccc} \hline Schnorr & ECDSA & EdDSA & RSA & Pairings \\\end{tabular} \footnote{2} & \begin{tabular}{cc} \hline NIST PQC \footnote{} & PQC \footnote{} \\\end{tabular} & \begin{tabular}{ccc} \hline Multi & Group & Ring \\\end{tabular} \footnote{} &  Applications \\
    \midrule

2023  \cite{buser2023survey} & \;\;\;\xmark & \begin{tabular}{p{0.7cm}p{0.5cm}p{0.5cm}p{0.5cm}p{0.3cm}} \hline \;\;\;\;\;\xmark & \xmark & \xmark & \xmark & \xmark \\\end{tabular} & \begin{tabular}{p{1cm}p{0.3cm}} \hline \;\;\;\;\;\xmark & \cmark \\\end{tabular} & \begin{tabular}{p{0.6cm}p{0.3cm}p{0.3cm}} \hline \;\;\;\cmark & \xmark & \cmark \\\end{tabular} & Blockchain\\ 

2023 \cite{csahin2023survey} & \;\;\;\xmark & \begin{tabular}{p{0.7cm}p{0.5cm}p{0.5cm}p{0.5cm}p{0.3cm}} \hline \;\;\;\;\;\xmark & \xmark & \xmark & \xmark & \xmark\\\end{tabular} & \begin{tabular}{p{1cm}p{0.3cm}} \hline \;\;\;\;\;\xmark & \cmark \\\end{tabular} & \begin{tabular}{p{0.6cm}p{0.3cm}p{0.3cm}} \hline \;\;\;\xmark & \cmark & \xmark \\\end{tabular} & \xmark \\

2023 \cite{ullah2023elliptic} & \;\;\;\xmark & \begin{tabular}{p{0.7cm}p{0.5cm}p{0.5cm}p{0.5cm}p{0.3cm}} \hline \;\;\;\;\;\xmark & \cmark & \xmark & \xmark & \xmark\\\end{tabular} & \begin{tabular}{p{1cm}p{0.3cm}} \hline \;\;\;\;\;\xmark & \xmark \\\end{tabular} & \begin{tabular}{p{0.6cm}p{0.3cm}p{0.3cm}} \hline \;\;\;\xmark & \xmark & \xmark \\\end{tabular} & \xmark\\ 

2022 \cite{perera2022survey} & \;\;\;\xmark & \begin{tabular}{p{0.7cm}p{0.5cm}p{0.5cm}p{0.5cm}p{0.3cm}} \hline \;\;\;\;\;\xmark & \xmark & \xmark & \xmark & \xmark\\\end{tabular} & \begin{tabular}{p{1cm}p{0.3cm}} \hline \;\;\;\;\;\xmark & \xmark \\\end{tabular} & \begin{tabular}{p{0.6cm}p{0.3cm}p{0.3cm}} \hline \;\;\;\xmark & \cmark & \cmark \\\end{tabular} & E-(voting, commerce, ... ) \\

2021 \cite{feng2022concretely} & \;\;\;\cmark & \begin{tabular}{p{0.7cm}p{0.5cm}p{0.5cm}p{0.5cm}p{0.3cm}} \hline \;\;\;\;\;\xmark & \xmark & \xmark & \xmark & \xmark \\\end{tabular} & \begin{tabular}{p{1cm}p{0.3cm}} \hline \;\;\;\;\;\xmark & \cmark \\\end{tabular} & \begin{tabular}{p{0.6cm}p{0.3cm}p{0.3cm}} \hline \;\;\;\xmark & \cmark & \xmark \\\end{tabular} & \xmark\\

2020 \cite{zhao2019secure} & \;\;\;\cmark & \begin{tabular}{p{0.7cm}p{0.5cm}p{0.5cm}p{0.5cm}p{0.3cm}} \hline \;\;\;\;\;\xmark & \xmark & \xmark & \xmark & \xmark \\\end{tabular} & \begin{tabular}{p{1cm}p{0.3cm}} \hline \;\;\;\;\;\xmark & \xmark \\\end{tabular} & \begin{tabular}{p{0.6cm}p{0.3cm}p{0.3cm}} \hline \;\;\;\xmark & \xmark & \xmark \\\end{tabular} & ML, Private set operations \\

2021 \cite{zhang2021survey} & \;\;\;\xmark & \begin{tabular}{p{0.7cm}p{0.5cm}p{0.5cm}p{0.5cm}p{0.3cm}} \hline \;\;\;\;\;\xmark & \xmark & \xmark & \xmark & \xmark \\\end{tabular} & \begin{tabular}{p{1cm}p{0.3cm}} \hline \;\;\;\;\;\xmark & \cmark \\\end{tabular} & \begin{tabular}{p{0.6cm}p{0.3cm}p{0.3cm}} \hline \;\;\;\xmark & \cmark & \xmark \\\end{tabular} & \xmark\\

2020 \cite{orsini2021efficient} & \;\;\;\cmark & \begin{tabular}{p{0.7cm}p{0.5cm}p{0.5cm}p{0.5cm}p{0.3cm}} \hline \;\;\;\;\;\xmark & \xmark & \xmark & \xmark & \xmark \\\end{tabular} & \begin{tabular}{p{1cm}p{0.3cm}} \hline \;\;\;\;\;\xmark & \xmark \\\end{tabular} & \begin{tabular}{p{0.6cm}p{0.3cm}p{0.3cm}} \hline \;\;\;\xmark & \xmark & \xmark \\\end{tabular} & \xmark \\

2020 \cite{aumasson2020survey} & \;\;\;\xmark & \begin{tabular}{p{0.7cm}p{0.5cm}p{0.5cm}p{0.5cm}p{0.3cm}} \hline \;\;\;\;\;\xmark & \cmark & \xmark & \xmark & \xmark \\\end{tabular} & \begin{tabular}{p{1cm}p{0.3cm}} \hline \;\;\;\;\;\xmark & \xmark \\\end{tabular} & \begin{tabular}{p{0.6cm}p{0.3cm}p{0.3cm}} \hline \;\;\;\xmark & \xmark & \xmark \\\end{tabular} & \xmark\\

2020 \cite{lindell2020secure} & \;\;\;\cmark & \begin{tabular}{p{0.7cm}p{0.5cm}p{0.5cm}p{0.5cm}p{0.3cm}} \hline \;\;\;\;\;\xmark & \xmark & \xmark & \xmark & \xmark \\\end{tabular} & \begin{tabular}{p{1cm}p{0.3cm}} \hline \;\;\;\;\;\xmark & \xmark \\\end{tabular} & \begin{tabular}{p{0.6cm}p{0.3cm}p{0.3cm}} \hline \;\;\;\xmark & \xmark & \xmark \\\end{tabular} & Privacy-preserving ML, ... \\

2020 \cite{fang2020digital} & \;\;\;\xmark & \begin{tabular}{p{0.7cm}p{0.5cm}p{0.5cm}p{0.5cm}p{0.3cm}} \hline \;\;\;\;\;\xmark & \cmark & \xmark & \xmark & \xmark \\\end{tabular} & \begin{tabular}{p{1cm}p{0.3cm}} \hline \;\;\;\;\;\xmark & \cmark \\\end{tabular} & \begin{tabular}{p{0.6cm}p{0.3cm}p{0.3cm}} \hline \;\;\;\xmark & \xmark & \xmark \\\end{tabular} & Blockchain\\ 

2020 \cite{ergezer2020survey} & \;\;\;\xmark & \begin{tabular}{p{0.7cm}p{0.5cm}p{0.5cm}p{0.5cm}p{0.3cm}} \hline \;\;\;\;\;\cmark & \cmark & \xmark & \xmark & \cmark \\\end{tabular} & \begin{tabular}{p{1cm}p{0.3cm}} \hline \;\;\;\;\;\xmark & \cmark \\\end{tabular} & \begin{tabular}{p{0.6cm}p{0.3cm}p{0.3cm}} \hline \;\;\;\xmark & \xmark & \xmark \\\end{tabular} & Certificate Authorities, Bitcoin\\


2019 \cite{cozzo2019sharing} & \;\;\;\cmark & \begin{tabular}{p{0.7cm}p{0.5cm}p{0.5cm}p{0.5cm}p{0.3cm}} \hline \;\;\;\;\;\xmark & \xmark & \xmark & \xmark & \xmark \\\end{tabular} & \begin{tabular}{p{1cm}p{0.3cm}} \hline \;\;\;\;\;\cmark & \xmark \\\end{tabular} & \begin{tabular}{p{0.6cm}p{0.3cm}p{0.3cm}} \hline \;\;\;\xmark & \xmark & \xmark \\\end{tabular} &  \\

2019 \cite{al2019efficient} & \;\;\;\xmark & \begin{tabular}{p{0.7cm}p{0.5cm}p{0.5cm}p{0.5cm}p{0.3cm}} \hline \;\;\;\;\;\xmark & \cmark & \xmark & \xmark & \xmark \\\end{tabular} & \begin{tabular}{p{1cm}p{0.3cm}} \hline \;\;\;\;\;\xmark & \xmark \\\end{tabular} & \begin{tabular}{p{0.6cm}p{0.3cm}p{0.3cm}} \hline \;\;\;\xmark & \xmark & \xmark \\\end{tabular} &  E-commerce \\

2019 \cite{al2019methods} & \;\;\;\xmark & \begin{tabular}{p{0.7cm}p{0.5cm}p{0.5cm}p{0.5cm}p{0.3cm}} \hline \;\;\;\;\;\xmark & \xmark & \xmark & \cmark & \xmark \\\end{tabular} & \begin{tabular}{p{1cm}p{0.3cm}} \hline \;\;\;\;\;\xmark & \xmark \\\end{tabular} & \begin{tabular}{p{0.6cm}p{0.3cm}p{0.3cm}} \hline \;\;\;\xmark & \xmark & \xmark \\\end{tabular} &  \xmark\\

2017 \cite{stathakopoulou2017threshold} & \;\;\;\xmark & \begin{tabular}{p{0.7cm}p{0.5cm}p{0.5cm}p{0.5cm}p{0.3cm}} \hline \;\;\;\;\;\xmark & \xmark & \xmark & \cmark & \cmark \\\end{tabular} & \begin{tabular}{p{1cm}p{0.3cm}} \hline \;\;\;\;\;\xmark & \xmark \\\end{tabular} & \begin{tabular}{p{0.6cm}p{0.3cm}p{0.3cm}} \hline \;\;\;\xmark & \xmark & \xmark \\\end{tabular} &  Blockchain \\

2016 \cite{mayer2016ecdsa} & \;\;\;\xmark & \begin{tabular}{p{0.7cm}p{0.5cm}p{0.5cm}p{0.5cm}p{0.3cm}} \hline \;\;\;\;\;\xmark & \cmark & \xmark & \xmark & \xmark \\\end{tabular} & \begin{tabular}{p{1cm}p{0.3cm}} \hline \;\;\;\;\;\xmark & \xmark \\\end{tabular} & \begin{tabular}{p{0.6cm}p{0.3cm}p{0.3cm}} \hline \;\;\;\xmark & \xmark & \xmark \\\end{tabular} &  Bitcoin and Ethereum \\

2016 \cite{subramanian3impact} & \;\;\;\xmark & \begin{tabular}{p{0.7cm}p{0.5cm}p{0.5cm}p{0.5cm}p{0.3cm}} \hline \;\;\;\;\;\xmark & \xmark & \xmark & \cmark & \xmark \\\end{tabular} & \begin{tabular}{p{1cm}p{0.3cm}} \hline \;\;\;\;\;\xmark & \xmark \\\end{tabular} & \begin{tabular}{p{0.6cm}p{0.3cm}p{0.3cm}} \hline \;\;\;\xmark & \xmark & \xmark \\\end{tabular} &  Cloud security \\

\midrule
\midrule
Ours  & \;\;\;\cmark & \begin{tabular}{p{0.7cm}p{0.5cm}p{0.5cm}p{0.5cm}p{0.3cm}}  \;\;\;\;\;\cmark & \cmark & \cmark & \cmark & \cmark \\\end{tabular} & \begin{tabular}{p{1cm}p{0.3cm}}  \;\;\;\;\;\cmark & \cmark \\\end{tabular} & \begin{tabular}{p{0.6cm}p{0.3cm}p{0.3cm}}  \;\;\;\cmark & \cmark & \cmark \\\end{tabular} & Blockchain, Cryptocurrency, IoT, Cloud, Healthcare, ...\\ 

\bottomrule
\end{tabular}

\begin{tablenotes}
\item 1 Including generic approaches using MPC protocols 
\inlineitem 2 Including conventional digital signatures standardized by NIST and approaches used for thresholding them
\inlineitem 3 Including thresholding post-quantum resistant digital signatures standardized by NIST
\inlineitem 4 Including thresholding non-standard post-quantum resistant digital signatures
\inlineitem 5 Including special signature schemes, namely group, ring, and multi-signatures
\end{tablenotes}

\end{table}

\subsection{The state-of-the-art Surveys and Limitations} \label{sec:StateofArtSurveys}
We have conducted a comprehensive and systematic investigation of recent studies focusing on threshold and special distributed signatures. The findings of our analysis are summarized in Table \ref{tab:publishedSurveys}. In our examination, we considered various aspects of these studies, including the \textbf{Approaches} adopted (generic or non-generic) (Column 2), the \textbf{Schemes} that were studied including conventional ones (Column 3) or NIST standardized and non-standard post-quantum signature schemes (Column 4), the \textbf{Special schemes} studied including the group, ring, and multi-signatures (Column 5), and the \textbf{Application} areas covered for the use of the studied signature(s) (Column 6).

{The initial observation in Table \ref{tab:publishedSurveys} highlights a notable gap in the existing literature regarding surveys that capture NIST (PQC) standards and exotic signatures through the lenses of threshold cryptography}. {\em (i)} The existing surveys discuss some distributed signatures mainly with custom thresholding (and in some rare cases with generic) approaches but do not offer an exhaustive examination from both custom and MPC-based generic methodologies with vis-a-vis comparison. {\em (ii)} This gap widens since existing work does not encompass custom and MPC-based thresholding of conventional and NIST-PQC standards as well as a wide variety of exotic features (e.g., \cite{chaum1991group,rivest2001leak,itakura1983public}). The importance of a wide range of thresholding capabilities is highlighted by NIST's recent call as the Multi-party Threshold Cryptography (MPTC) project \cite{brandao2023nist}. {\em (iii)} Finally, the scope of application domains regarding threshold signatures in existing surveys usually are limited to only a few domains (e.g., Blockchains). Below, we discuss some of these gaps in further detail.

\vspace{-2mm}
\paragraph{\underline{Limited Coverage on Thresholding Approaches and Their Comparison}}: The custom-design of distributed signatures such as multi-signatures~\cite{liu2023idenmultisig}, group signatures~\cite{boneh2004group}, and some custom-design threshold signatures~\cite{komlo2021frost} received attention in the literature. Yet, another angle to threshold digital signature is the use of MPC~\cite{cozzo2019sharing}. Although MPC and some of its applications have been surveyed in the literature~\cite{lindell2020secure,orsini2021efficient,zhao2019secure,feng2022concretely}, only little attention (e.g., \cite{buser2023survey}) has been given to transparent thresholding of signatures with MPC~\cite{cozzo2019sharing,brandao2023nist}. In particular, there is a gap in encompassing transparent thresholding of NIST-PQC standards and exotic signatures (conventional or PQ-safe).    

\vspace{-2mm}
\paragraph{\underline{Narrow/Selective Focus on Conventional-custom and Exotic Distributed Signatures}} Existing studies mainly focus on only some groups of signatures (e.g., some EC-based signatures~\cite{ullah2023elliptic}, ECDSA~\cite{aumasson2020survey,al2019efficient}). However, for example, thresholding the other digital signatures like BLS with applications in IoT \cite{grissa2019trustsas} and Blockchain \cite{santiago2021concordia},  Schnorr in Blockchains or Cryptocurrency \cite{nick2021musig2,nick2020musig} holds equal importance. In the same line, although there have been commendable efforts in surveying exotic signatures (e.g., \cite{buser2023survey,csahin2023survey,perera2022survey,feng2022concretely,zhang2021survey}), they usually focus on certain types of application domains (e.g., Blockchain, IoT), and capture some exotic features in isolation. Furthermore, the examination of threshold PQC signatures, encompassing both the NIST-PQC standards and custom design~\cite{bonte2021thresholdizing}, is another gap that has not been addressed fully. While certain studies \cite{buser2023survey,cozzo2019sharing} have investigated non-standard threshold signatures for Blockchain and NIST-PQC signatures, respectively, there is a noticeable gap in capturing standardized, custom, and exotic signatures simultaneously, especially with MPC in mind (as also observable from Table \ref{tab:publishedSurveys}).

\vspace{-2mm}
\paragraph{\underline{Limited Attention to Broad Applications and Future Directions}}: The practical applications of threshold signatures have significant importance, yet we observed that only a limited number of papers have examined applications of threshold signatures, primarily centered around Blockchain (e.g.,~\cite{buser2023survey}) and e-voting/currency (e.g., with PQC~\cite{csahin2023survey}) domains. Furthermore, it is ideal to consider visions towards the future evolution of threshold digital signatures. However, we observe that this important angle received less attention in the majority of the current surveys. Hence, there is a need for forward-looking perspectives in the design and applications of emerging threshold digital signatures.

%% file: bodies/contribution.tex
\begin{figure}[ht]
    \centering
    \includegraphics[width=\linewidth]{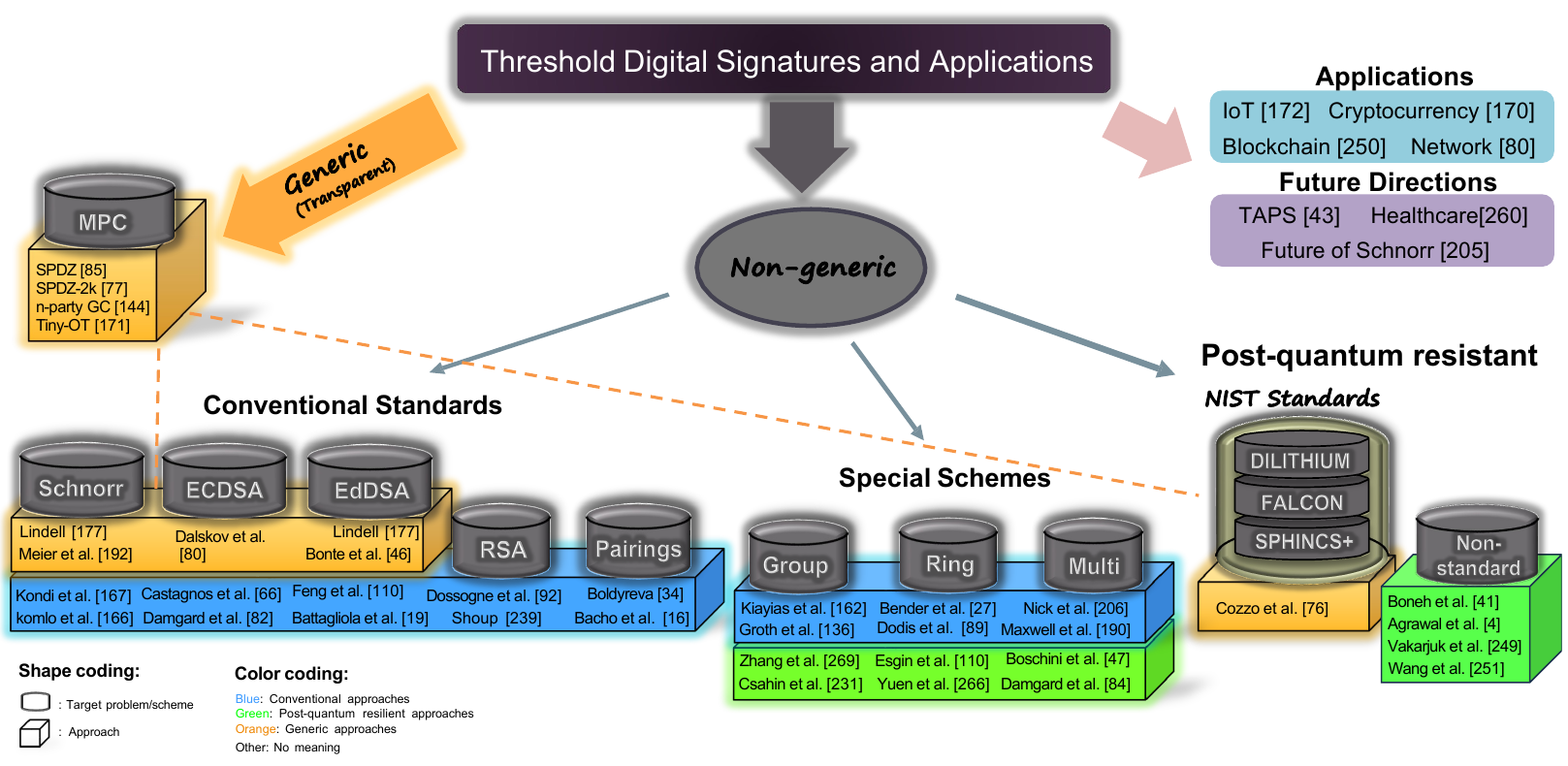}
    \caption{High-level Taxonomy of Our Survey.}
    \label{fig:taxonomy}
\end{figure}

This paper aims to fill the aforementioned gaps in the state-of-the-art by presenting a comprehensive and systematic survey on threshold digital signatures. To the best of our knowledge, this is the first survey that encompasses threshold digital signatures both in conventional and PQC settings, with NIST (PQC) standards and custom designs via generic (MPC) and non-generic thresholding mechanisms, while also capturing exotic signatures and real-life applications. The taxonomy of our survey is depicted in Figure \ref{fig:taxonomy}, and our contributions are further outlined as follows: 

\vspace{-1mm}
\paragraph{\underline{Extensive Coverage of Custom and Generic Thresholding with a Vast Range of Digital Signatures}}: We examine both custom-designed and generic thresholding mechanisms applied to a myriad of digital signature families to offer broad coverage of distributed digital signatures. Our survey places a special emphasis on MPC-based thresholding, and in line with NIST's recent efforts~\cite{brandao2023nist}, we posit that such generic thresholding will play a crucial role in the future threshold digital signatures. That is, the MPC-based approach offers \textit{transparent thresholding} that respects the algorithmic flow and features of the base scheme. Hence, MPC-based thresholding preserves the provable security, signature size,  and verification of the base scheme with full-standard compliance, which is an important requirement for some real-life applications. In this direction, in Section~\ref{sec:genericmpc}, we introduce generic thresholding with MPC, capture the works that MPC thresholds prominent conventional-secure primitives (e.g., MPC-Schnorr~\cite{meier2022mpc}, MPC-ECDSA~\cite{ECDSAANSI} and MPC-EdDSA~\cite{bonte2021thresholdizing}), and finally NIST-PQC signatures~\cite{nistSelectedAlgorithms} in Section~\ref{sec:PQC}. We present MPC gadgets and capture the relationship between custom and generic thresholding approaches, especially in PQC settings, with detailed comparisons in our tables.

Despite its merits, generic thresholding is usually (significantly) costlier than custom-designed threshold signatures, which constitute the vast majority of the current threshold signature landscapes. Accordingly, our survey gives a prime coverage of such threshold signatures, as further elaborated below.

\vspace{-2mm}
\paragraph{\underline{Comprehensive Examination of Threshold Conventional/PQ-secure Signatures and Standards}}: We offer a comprehensive examination of custom-designed threshold digital signatures in broad classes of signature families based on \textit{Discrete Logarithm Problem (DLP)}, \textit{Elliptic Curve Discrete Logarithm Problem (ECDLP)}, \textit{Cryptographic Pairing}, \textit{Integer Factorization (IF)}, and \textit{Lattices}. (i)  Our examination begins in Section \ref{sec:conventionalSchnorr} with a focus on the Schnorr-like signatures and relevant standards such as ECDSA and EdDSA due to their prevalence in real-life applications and efficiency advantages. Our analysis goes beyond a mere summary by highlighting algorithmic relations and the pros and cons of each technique when threshold settings are considered. (iii) We then continue analyzing BLS-based and RSA-based threshold signatures in Section \ref{sec:pairing} and Section \ref{sec:rsa}, respectively, which capture cryptographic pairing and IF-based approaches as important classes of digital signatures. As discussed before, we also cover threshold PQC signatures (e.g., lattice-based) in Section \ref{sec:PQC}, both with custom and MPC-based thresholding techniques. (iv) In each category, we feature detailed analysis via tables that offer a chronological overview of notable previous works, providing succinct yet comprehensive explanations of the methodologies employed, accompanied by performance assessments (when applicable).

\vspace{-2mm}
\paragraph{\underline{Investigation of Exotic Features and Real-life Applications}} In the context of other distributed signatures with exotic features, in Section \ref{sec:specialschemes}, we first offer a succinct analysis of the main constructions within each scheme and then their relationship with threshold settings. We focus on Group \cite{zhou2006shorter} (Section 8.1), Ring \cite{bender2006ring} (Section 8.2), and Multi-signatures \cite{nick2021musig2} (Section 8.3) in conventional and PQ-secure settings. Finally, in Section \ref{sec:applications}, we examine essential applications of threshold and exotic signatures in emerging real-life applications and also briefly discuss potential future directions such as NIST's call for additional PQC signatures \cite{nistStandardizationAdditional} and others.

%% file: bodies/notations.tex
\subsection{Notations}

$\khash$ denotes a secure cryptographic hash function (e.g., SHA-256). $\knorm$ denotes the infinity norm. $\kstrcat$ denotes string concatenation. $\kpprime$ and $\kqprime$ represent large prime numbers. $\kmessageanylen$ represents a message of any finite length. $\kprivkey$ and $\kpubkey$ represent private and public keys, respectively. $\kgroup$ and $\kcycgroupgen$ represent a cyclic group with its generator, respectively. $\keccurve$, $\keccurveident$, $\keccurvebase$, and $\kecorder$ represent an elliptic curve, the identity element, the base point and its order, respectively. $\kecmult$ represents elliptic curve point multiplication by a scalar. Non-bold font $x$ denotes a scaler, and bold $\textbf{x}$ denotes a vector. $\kcyczoverp$ denotes a finite cyclic group of order $\kpprime-1$. $\mathbb{F}_p$ denotes a finite field of order $\kpprime-1$. We define the other notation and variables for specific schemes and definitions in place.

\subsection{Intractability Assumptions and Mathematical Backgrounds}
We concisely overview the intractability assumptions and definitions used in this work.

\kdefrsa
\kdefdlp
\kdefecdlp

\kdefddhp
\kdeflattice
\kdefsis
\kdeflwe
\kdefmlwe
\kdefmsis

\textit{Secret sharing} is a technique used in \textit{Threshold cryptography}, in which a threshold scheme involves a dealer who holds a secret, a group of $n$ parties known as shareholders, and a collection of subsets of parties called the \textit{Access structure}. The secret is divided among the shareholders using a deterministic method that ensures each share reveals no information about the original secret. However, any set in the access structure can reconstruct the secret. Secret sharing is used in various applications of cryptography and distributed computing, such as Byzantine agreement \cite{rabin1983randomized}, Digital signatures \cite{desmedt2001shared}, Secure multi-party computation \cite{cramer2000general}, Attribute-based encryption \cite{waters2011ciphertext}, etc. Shamir \cite{shamir1979share} and Blakely \cite{blakley1979safeguarding} introduced the first secret sharing schemes. Shamir's scheme is more practical today and is based on \textit{Lagrange polynomial interpolation} \cite{werner1984polynomial} while Blakely's scheme is based on \textit{Solving linear systems}. 

\kdefshamirsss

%% file: bodies/digital_signatures.tex
\subsection{Digital Signatures}

Digital signatures \cite{chen2023digital} are essential cryptographic mechanisms that enable scalable authentication with non-repudiation and public verifiability through the use of \textit{Public Key Infrastructure (PKI)}. In digital signatures, a signer $S$ has a key pair $(\kprivkey, \kpubkey)$ and signs a message using the private key. Anyone possessing the signer's public key can verify the integrity of the message and ensure that it has remained unaltered during transmission.

Various digital signature models have been developed for different requirements \cite{cao2008classification}. In this study, we will specifically focus on \textit{Threshold digital signatures} \cite{desmedt2001shared}, \textit{Multi-signatures} \cite{itakura1983public}, \textit{Group signatures} \cite{chaum1991group}, and \textit{Ring signatures} \cite{rivest2001leak}. It is worth mentioning that the latter three models are considered to be distributed signature schemes and are referred to as \textit{Exotic (or Special)} schemes within our work. Each model will be investigated in detail in its corresponding section.

\kdefdigitalsignature

\subsubsection{Security Models of Digital Signatures} The security model used for digital signatures is \keucmadesc.

\kdefeucma

%% file: bodies/threshold_digital_signatures.tex
\subsection{Threshold Digital Signatures} \label{sec:ThresholdDS}

 \kdefthresholds

\subsubsection{Security Models of Threshold Digital Signatures} 
Two security definitions can be used for threshold digital signatures. The $\ktprivate$ notion is general to any thresholded scheme, and the 
$\kteucma$ is the thresholded definition of $\keucmanotion$ mentioned in Definition \ref{def:eucma}.

\kdeftprivateprot
\kdefteucma

%% file: bodies/MPC.tex
\subsection{Multi-party Computation (MPC)}\label{sec:mpcdef}
\textit{(Secure) Multi-party Computation (MPC)} allows a group of distributed computing parties to jointly compute a function using their private inputs, ensuring that no additional information is disclosed during the computation process except for the function's output. The concept of MPC was initially introduced by Andrew Yao \cite{yao1982protocols} as the "The Millionaires" problem for two parties and later extended to multiple parties by Goldreich et al. \cite{goldreich2019play}. Later, in his subsequent work \cite{yao1986generate}, he presented the \textit{Garbled circuits} protocol as a two-party MPC protocol with a fixed number of rounds involving one party garbling the target function, initially represented as a circuit of logical gates, and sending it to the other party for final computation. Throughout this process, additional communications are needed to assist the other party in computing the function, which is facilitated by techniques like \textit{Oblivious transfer} \cite{halevi2012smooth}. Following Yao's protocol, many other protocols in the literature, such as GMW \cite{goldreich2019play}, BGW \cite{ben2019completeness}, BMR \cite{beaver1990round}, and GESS \cite{kolesnikov2005gate} have been proposed.

Initially, research on MPC primarily concentrated on theoretical aspects, aiming to define and prove the security of presented protocols under different adversarial and network models using complex operations that resulted in excessive overhead. However, numerous early works have contributed to improving the efficiency of MPC constructions via computation algorithm optimizations, memory and communication overhead improvements, and leveraging CPU-level instructions like AES-NI. Subsequent attempts optimized the circuit computation cost. Unlike AND/OR gates, XOR gates are computationally inexpensive \cite{kolesnikov2008improved}, and using them in the circuit can significantly improve the performance. While constructing the circuits manually \cite{huang2012private} was challenging, researchers decided to use \textit{Special-purpose MPC Compilers} \cite{hastings2019sok} to generate efficient circuits for the functions written in the \textit{Secure Function Definition Language (SFDL)} language.

The security evaluation of MPC protocols involves considering factors such as the adversary's behavior, corruption strategy, the number of compromised parties, and the execution environment:

\underline{Adversary's behavior :} A \textbf{Semi-honest (passive)} adversary is an attacker who corrupts parties and acquires their internal state, including private input and message transcripts, while adhering to the protocol. In contrast, a \textbf{Malicious (active)} adversary may force corrupted parties to deviate from the protocol. An adversary is considered \textbf{Covert} if they attempt to breach the protocol. This security breach, however, can be detected with a specific probability and can be further adjusted accordingly.

\underline{Parties' corruption:} \textbf{Static} corruption refers to a scenario in which honest parties maintain their honesty throughout the entire protocol execution, and the corrupted parties remain unchanged for the whole duration of the protocol execution. \textbf{Adaptive} corruption, on the other hand, occurs when the adversary, based on the knowledge acquired, selects which parties to corrupt. \textbf{Proactive} corruption involves a situation where certain parties can be corrupted while the protocol progresses. However, these corrupted parties remain compromised for a specified duration and return to an honest state once a security breach is detected.

\underline{Number of corrupted parties:} When the adversary's ability to corrupt is limited to fewer than half of the parties, the security model is characterized as \textbf{honest majority}. Conversely, if the adversary can corrupt more than or equal to half of the parties, the model is \textbf{dishonest majority}.

\underline{Execution environment:} A framework proposed by Canetti \cite{canetti2001universally} augments the security models with the consideration of the environment. This framework facilitates the proof of the protocol's security independent of other concurrently running protocols within the environment.

%% file: bodies/conventional_approaches.tex
\section{Threshold Schnorr-like Digital Signatures and Standards} \label{sec:schnorr}

The NIST \cite{chen2023digital} has standardized digital signatures that rely on three well-known security problems: \textit{Integer Factorization (IF)}, \textit{Discrete Logarithm Problem (DLP)}, and \textit{Elliptic Curve Discrete Logarithm Problem (ECDLP)} (See \ref{def:IntFactP}, \ref{def:dlp}, and \ref{def:ecdlp}). In the upcoming sections, we will explore thresholding of \textit{Schnorr} signature scheme \cite{brandao2023notes} based on the DLP and its (elliptic curve)-based variants including \textit{EdDSA} \cite{josefsson2017edwards,brandao2023notes} and \textit{ECDSA} \cite{chen2023digital} based on the ECDLP.

\subsection{Threshold Schnorr-based Signatures} \label{sec:conventionalSchnorr}
\begin{figure}
\centering

\vspace{-5mm}
\caption*{(a) (DLP and ECDLP)-based Schnorr Signature Comparison \cite{brandao2023notes}}
\vspace{-1mm}
\includegraphics[width=120mm, height=35mm]{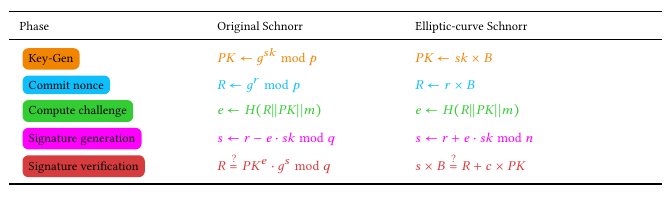}

\vspace{-3mm}
\caption*{(b) Variants of Schnorr Signature Scheme}
\vspace{-3mm}
\begin{multicols}{3}

\includegraphics[width=51mm]{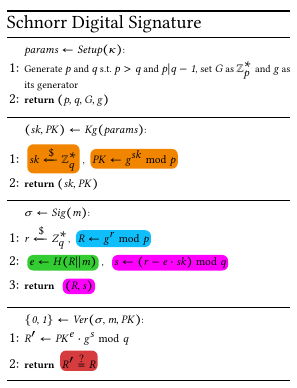}

\columnbreak

\includegraphics[width=51mm]{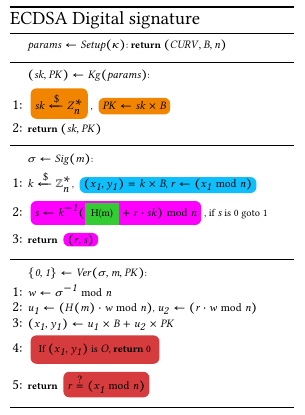}

\columnbreak

\includegraphics[width=51mm]{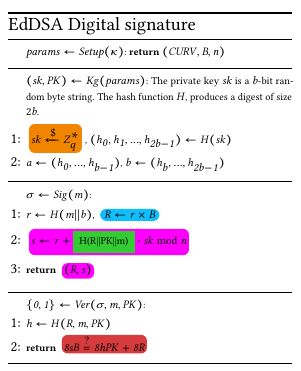}
\end{multicols}

\vspace{-6mm}
\caption*{(c) FROST Signature Scheme \cite{komlo2021frost}}

\vspace{-3mm}
\begin{multicols}{2}

\includegraphics[scale=0.43]{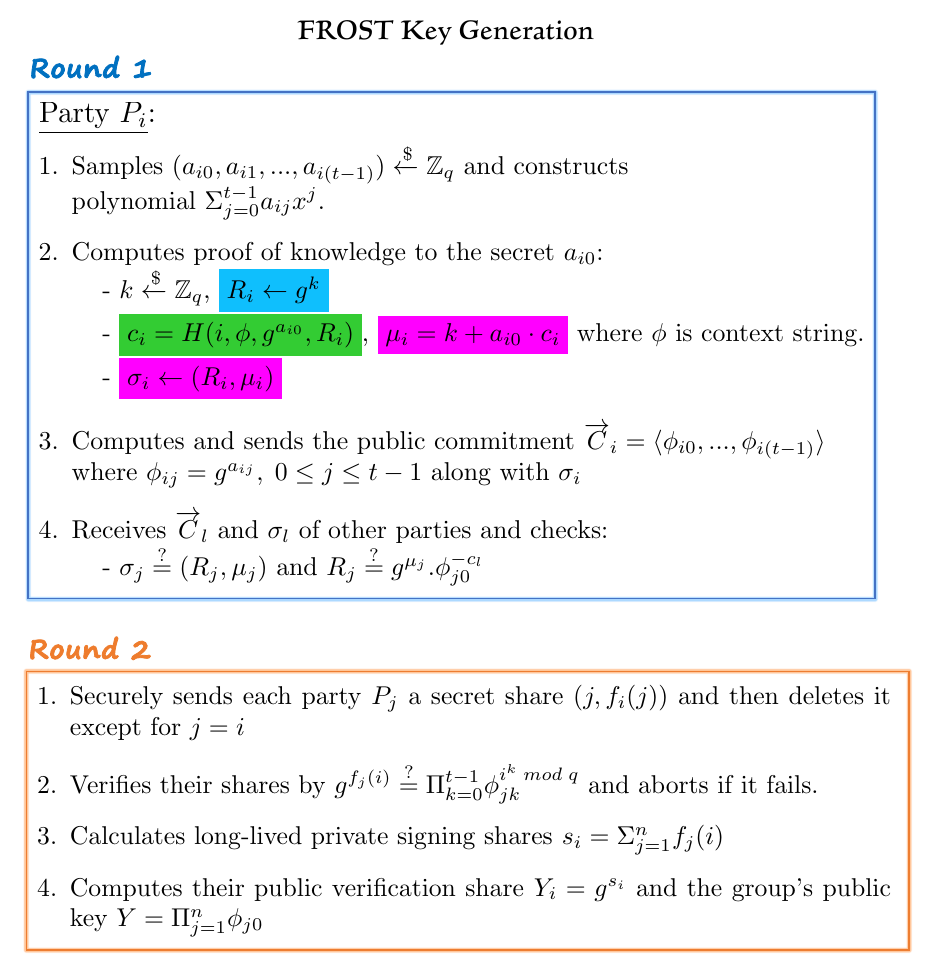}

\columnbreak

\includegraphics[scale=0.43]{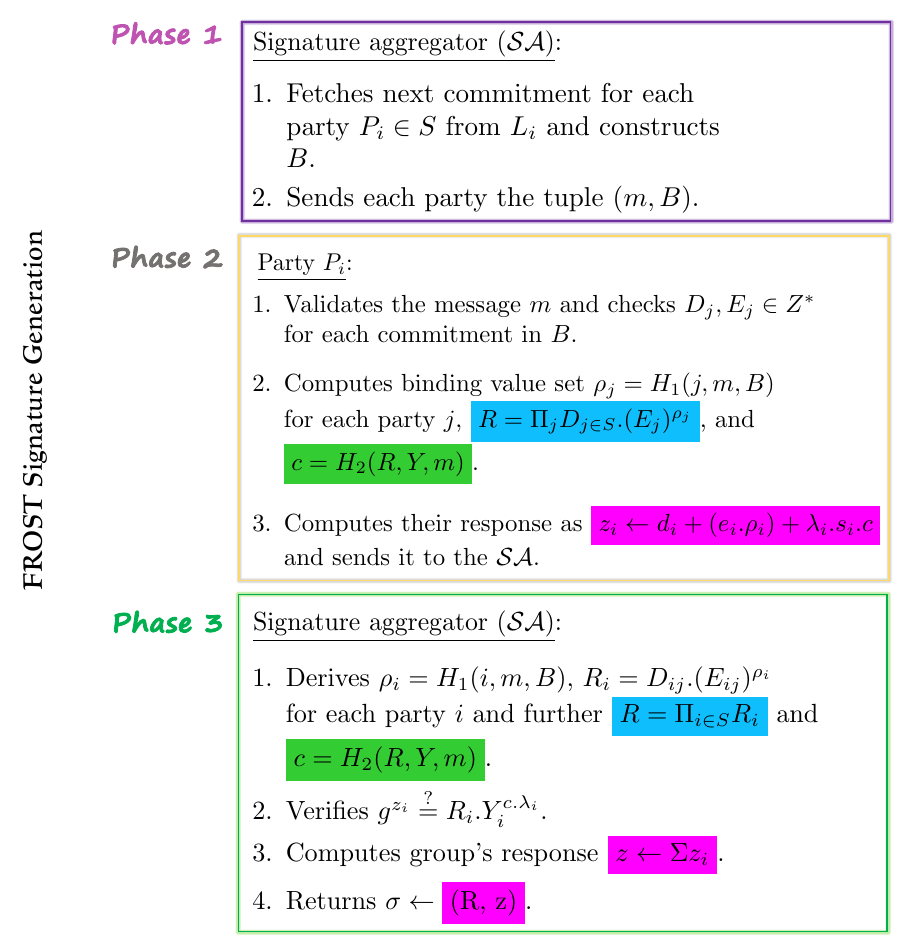}

\end{multicols}

\vspace{-4mm}
\caption{Schnorr Signature and its Variants}
\label{fig:SchnorrItsVariants}
\end{figure}

The Schnorr signature, introduced by Claus Schnorr \cite{schnorr1990efficient}, is a DLP-based digital signature. Various existing standards, such as ECDSA and EdDSA, are EC-based variants of the Schnorr signature. A seminal work of Horster et al. \cite{horster1994meta} integrated a broad class of approaches regarding Schnorr-like and Elgamal-like signatures and presented a general \textit{Meta-ElGamal} family. Inspired by this Meta-Elgamal family, we present Figure \ref{fig:SchnorrItsVariants} to illustrate the relationship between Schnorr-like signatures. Specifically, the comparison of the operations of the original Schnorr scheme operating on integer groups and EC-Schnorr variants along with their algorithms have been illustrated in Figures \ref{fig:SchnorrItsVariants}(a) and \ref{fig:SchnorrItsVariants}(b), respectively. Table \ref{tab:ThreshooldSchnorr} contains a compilation of recent works on threshold Schnorr signatures.

Benhamouda et al. \cite{benhamouda2023sprint} proposed a recent work on threshold Schnorr signatures. They focused on developing high-throughput protocols that enable parallel signing of messages in a single run. These protocols were designed for both synchronous and asynchronous with special attention on distributed $(t,n)$ Schnorr signatures on asynchronous public Blockchains. They addressed the challenge of computing the value $R$ by using \textit{Distributed Key Generation (DKG)} \cite{gennaro2007secure} to secretly share the nonce $r$ among the parties. Unlike traditional DKG schemes that require \textit{Complete secret-sharing}, their approach allows sufficient honest parties possessing the necessary secrets to proceed with the protocol and finish it before sharing is complete. They utilized \textit{Packed secret sharing} \cite{franklin1992communication} and \textit{Super-invertible matrices} \cite{hirt2006robust} to speed up the generation of ephemeral secrets needed for DKG runs, although it comes at the cost of reduced resiliency in the parameter $t$. Moreover, instead of relying on existing MPC solutions that involve interactions, they proposed sharing packs of long-term secrets instead of individual values. This eliminates interactions in the signature generation process, enabling parties to perform SIMD partial signature generation through local multiplications. Further optimization techniques were incorporated, including a modular QUAL agreement protocol over broadcast channels and a randomness beacon for implementing sub-sampling for achieving smaller sub-samples committees. These techniques are particularly advantageous in environments with a large number of parties. In terms of performance, their scheme can generate $\Omega(n^2)$ signatures in each run and requires $O(n^2)$ broadcast bandwidth, where $n$ is the size of the committees.

Work by Komlo et al. \cite{komlo2021frost} proposed a round-optimized threshold Schnorr signature known as FROST with signature operations depicted in Figure \ref{fig:SchnorrItsVariants}(c). Many threshold signatures prioritize robustness over efficiency, in which the honest parties complete the protocol after finding and eliminating the misbehaving party. However, FROST is more optimistic regarding finding malicious parties and tries not to sacrifice efficiency, and parties will abort the protocol and rerun it if they detect a misbehaving party. Another optimization taken into account by FROST is the role of a semi-trusted signature aggregator ($\mathcal{SA}$) in the scheme to reduce the communication overhead. This party can either be played by one of the parties or by an external party. Compromising the $\mathcal{SA}$ won't let the adversary learn the private key or forge the signature. Regarding the key generation phase, FROST uses a variant of Pedersen's DKG proposed in \cite{gennaro2003secure} in which via using Feldman's VSS \cite{feldman1987practical}, each party gets their secret share as the sum of the shares received from all the VSS executions. Furthermore, each party has to prove their secret in zero-knowledge to other parties, which can protect against rogue-key attacks \cite{bellare2002randomness} when $t \ge  n/2$. The signature generation phase uses the simple additive secret sharing and share construction techniques used by other methods. Signature generation consists of two phases: The first phase is the preprocessing phase, which aims to enhance the scheme's security against the Drijvers et al. \cite{drijvers2019security} attacks by establishing a robust connection between each party's response with a specific message and a designated group of committed parties that are related to that signing operation. In this phase, nonces can be generated non-interactively in a distributed manner. This phase ensures that signature generation is efficient and secure. The second phase is the actual signing phase depicted in Figure \ref{fig:SchnorrItsVariants}(c)(right). Note that the signature generation phase can be seamlessly integrated with the preprocessing phase, resulting in a unified two-round protocol.

\begin{table}[H]
\scriptsize
\centering
  \caption{Summary of Threshold Schnorr Digital Signatures}
  \label{tab:ThreshooldSchnorr}
  \begin{tabular}{p{1cm} p{0.9\linewidth}}
  \toprule
Work by & Description\\ 
\midrule
2023 \cite{benhamouda2023sprint} & (See \ref{sec:conventionalSchnorr} for details)\\ 
\midrule 

2023 \cite{crites2023fully} & \ding{50} Proposing a three-round threshold Schnorr signature that, unlike many schemes secure against the static adversary, achieves adaptive security \ding{50} By assuming $\mathit{t+1}$ as the threshold, the proposed scheme achieves static security with $\mathit{t}$ corruption under the DLP and ROM assumptions, adaptive security with $\mathit{t/2}$ corruptions under the \textit{Algebraic One-More Discrete Logarithm Assumption (AOMDL)} and ROM assumptions, and full adaptive security with $\mathit{t}$ corruptions under the \textit{Algebraic Group Model (AGM)}, AOMDL, and DLP assumptions $\star$ The variant with static security is a 3-round protocol which compared to the work \cite{komlo2021frost} requires fewer exponentiations but one more hash operation\\
\midrule

2023 \cite{kondi2023two} & \ding{50} Proposing an approach for distributing Schnorr signature scheme with stateless signing procedure based on using \textit{Pseudorandom Correlation Functions (PCFs)} \cite{boyle2020correlated}, multi-party analog of PRFs, which make the deranomization technique a distributed procedure \ding{50} Giving two instantiations of the proposed approach based on \textit{Vector Oblivious Linear Evaluation (VOLE) correlation}: First scheme is a two-round protocol based on \cite{roy2022softspokenot} which achieves covert security and uses PRF and OT as the assumptions. The second scheme is a two-round protocol based on the Paillier encryption-based PCF of \cite{orlandi2021rise} that achieves full malicious security and uses \textit{Decisional Composite Residuosity (DCR)} and \textit{Strong RSA} as the assumptions $\star$ Achieving 0.1KB of bandwidth overhead in the first instantiation and 0.5KB in the second one $\star$ Mentioning achieving both covert and malicious security simultaneously as an open problem\\
\midrule

2022 \cite{joshi2022atssia} & \ding{50} Proposing a non-interactive asynchronous threshold schnorr signature \ding{50} Improving the scheme \cite{komlo2021frost} by having the signing operation running independent of the parties and not requiring the signers to be online all in once until the aggregation phase \ding{50} Compared to \cite{komlo2021frost}, it is robust, does not have a preprocessing stage and allows dynamic participation \ding{50} Secure against ROS solver \cite{benhamouda2022security} and Drijvers et al. \cite{drijvers2019security} attacks in ROM  \\
\midrule

2022 \cite{ruffing2022roast} & \ding{50} Proposing a wrapper that takes in a semi-interactive threshold signature scheme (i.e., signing consists of preprocessing and signing rounds) and converts it to a protocol that is robust and asynchronous \ding{50} Applying the wrapper to the FROST scheme and obtaining an efficient and robust Schnorr signature scheme \ding{50} Providing robustness through having concurrent signing sessions and reduce the exponential number of the sessions from $\mathit{\kcombc{t}{n}}$ to at most $\mathit{n-t+1}$ $\star$ Using FROST for benchmark with (67, 100) threshold setting running in 3 seconds$\star$ Possible improvements by supporting a group of multiple input messages simultaneously, allowing continuous stream signing, and letting signers be in more than one session to reduce latency\\
\midrule

2022 \cite{battagliola2022provably} & \ding{50} Proposing a (2, 3) threshold Schnorr-based variant of \cite{battagliola2022threshold} with an offline recovery participant for crypto-assets custody \ding{50} Provable security against adaptive adversaries \ding{50} Capable of achieving resiliency of recovery in the presence of a malicious adversary if the encryption scheme supports DLP verification using ZKP \ding{50} Using an IND-CPA \cite{marcedone2014obfuscation} encryption scheme for communicating with the offline party along with a secure non-malleable commitment scheme  \ding{50} Using ZKP of Knowledge (ZKPoK) to prove the binding of the secret values and their corresponding public values \ding{50} Using Feldman's VSS \cite{feldman1987practical} for the secret sharing \\
\midrule

2021 \cite{komlo2021frost} & (See \ref{sec:conventionalSchnorr} for details) \\
\bottomrule
\end{tabular}
\ktablehelp
\end{table}
















\subsection{Threshold ECDSA} \label{sec:ecdsa}
The Elliptic-curve Digital Signature Algorithm (ECDSA), illustrated in Figure \ref{fig:SchnorrItsVariants}(b), is a variant of the Schnorr signature that utilizes elliptic curves and has been specified by ANSI \cite{ECDSAANSI} and NIST \cite{chen2023digital}. One of its approved variants specified by Pornin \cite{pornin2013deterministic} enables deterministic signature generation. ECDSA is widely used in different applications, including secure DNS (DNSSEC) \cite{dalskov2020securing}, TLS, and facilitating cryptocurrencies like Bitcoin \cite{goldfeder2015securing}. Due to this versatile application, unlike other schemes, there have been significant efforts in proposing two-party or multi-party threshold ECDSA signatures where they aim to improve existing schemes in terms of communication rounds, overhead, signing time, and introduce additional properties such as trustless setup, identifiable aborts, etc. Table \ref{tab:tableECDSAApproaches} contains a compilation of recent fundamental works focusing on threshold ECDSA signatures. 

To establish a threshold ECDSA scheme, two essential components need to be distributed: the private key $sk$ and the nonce $k$. However, distributing these components poses challenges when computing the nonce commitment $R$ and multiplying $k$ and $sk$ in the signature generation process to obtain the signature portion $s$. Earlier schemes, including the one proposed by Gennaro et al. \cite{gennaro2016threshold}, tackled the challenge by relying on Paillier's cryptosystem. However, these schemes had limitations, such as the need for lengthy messages and complex and computationally intensive zero-knowledge proofs. Gennaro et al. \cite{gennaro2018fast} introduced a more efficient approach by incorporating the SPDZ technique developed by Damgård et al. \cite{damgaard2012implementing} and employed a two-party protocol initially proposed by Gilboa et al. \cite{gilboa1999two}, wherein parties worked in pairs and transformed their multiplicative shares into additive shares. This protocol enabled the combination of two shared secrets, denoted as $(x_1,x_2,...,x_n)$ and $(y_1,y_2,...,y_n)$, resulting in $\Sigma_{i,j}{x_i\cdot y_j}$. This improved approach offered a streamlined solution to address the aforementioned challenge.

Castagnos et al. \cite{castagnos2020bandwidth} proposed an alternative method to improve the previously proposed scheme in \cite{gennaro2016threshold}. They identified challenges associated with using Paillier's cryptosystem in the scheme, particularly when dealing with malicious adversaries. The mismatch between Paillier plaintexts residing in $(\mathbb{Z}/N\mathbb{Z})$ and ECDSA signatures residing in $(\mathbb{Z}/q\mathbb{Z})$ necessitated selecting a large value for $N$ to prevent wraparounds, which resulted in increased complexity. Additionally, range proofs were required to verify the range of encrypted values, adding to the scheme's complexity. To overcome these challenges associated with the Paillier cryptosystem, Castagnos et al. introduced their previously proposed technique \cite{castagnos2019two}, which utilized hash-proof systems to construct a two-party ECDSA scheme. This technique enabled the instantiation of class groups of imaginary quadratic fields using Castagnos and Laguillaumie's homomorphic encryption scheme \cite{castagnos2015linearly}. By leveraging this approach, the message space could be defined as $(\mathbb{Z}/q\mathbb{Z})$. Modifications were also implemented to address weaknesses in the original encryption scheme, ensuring that ciphertexts were well-formed and avoiding issues of non-surjectivity.

Castagnos et al. \cite{castagnos2023bandwidth} extended their previous work \cite{castagnos2020bandwidth} to develop a threshold ECDSA signature that incorporates accountability and online/offline efficiency with provable security. They also extended the scheme to support adaptive security in the $(n, n)$ setting. Building upon the concept of online/offline computation introduced by Gennaro et al. \cite{gennaro2020one} (superseded by Canetti et al. \cite{canetti2020uc}), they designed a scheme where the shared computation of $k^{-1}\cdot sk$ and $k$ takes place in the offline stage. In the online phase, individual parties compute their signature shares as $s_i=k^{-1}_i(H(m)+r\cdot sk_i)$ non-interactively, which are then combined to produce the final signature $s$. To detect malicious parties, zero-knowledge proofs can be utilized, but their integration into the protocol can introduce significant complexity. Instead, the approach proposed by \cite{gennaro2020one} allows the detection of malicious parties by verifying the public value $s_i$ computed by each party in the online phase, as well as the random choices revealed by each party in the offline phase. However, this approach cannot be directly applied to the scheme proposed in \cite{castagnos2019two} due to the use of hash proof systems, which prevent the completion of the proof. To address this limitation, they introduced new zero-knowledge proofs that enable the simulator to complete the proof without compromising overall efficiency.

\begin{table}[H]
\scriptsize
\centering
  \caption{Summary of Threshold ECDSA Digital Signatures}
  \label{tab:tableECDSAApproaches}
  \begin{tabular}{p{1cm} p{0.9\linewidth}}
  \toprule
Work by & Description\\ 
\midrule
2023 \cite{castagnos2023bandwidth} &  (See \ref{sec:ecdsa} for details) \\
\midrule

2022 \cite{battagliola2022threshold} & \ding{50} Proposing the first $(2, 3)$ threshold ECDSA scheme with allowing the presence of one offline party for signature recovery \ding{50} With four possible phases: \textit{Setup} phase played by all parties exchanging public parameters, \textit{Key generation} played by two players only with third one already chose a long-term asymmetric key and went offline, \textit{Signing} Played by two players, and \textit{Signature recovery} Played with the third player and one of the other ones \ding{50} Proven security against adaptive security based on standard assumptions \ding{50} Requiring provable encryption scheme instead of a CPA asymmetric encryption to deal with malicious offline party\\
\midrule

2022 \cite{damgaard2022fast} & \ding{50} Proposing a bandwidth efficient four-round ($\mathit{t, n}$) threshold ECDSA scheme with two phases of preprocessing (three rounds) and signing phase (one round) \ding{50} Proving secure against active adversary in the honest majority model for $\mathit{n \ge 2t + 1}$ and with a UC security proof \ding{50} Using Shamir's secret sharing over Pedersen's verifiable secret sharing for joint random secret sharing of the private key \ding{50} Extending the proposed method to support fairness and termination $\star$ Achieving $\mathit{n\log{}q}$ bits communication with two long curve multiplication in signing, $\mathit{6n \log{}q + 2n \log{}p}$ bits communication with three long curve multiplication in presigining and $\mathit{n \log{} q + n \log{} p}$ bits communication in the key generation phase\\
\midrule

2020 \cite{castagnos2020bandwidth} & (See \ref{sec:ecdsa} for details)  \\

\bottomrule
\end{tabular}
\ktablehelp
\end{table}

\begin{table}[H]
\scriptsize
\centering
  \caption{Summary of Threshold ECDSA Digital Signatures (Cont.)}
  \label{tab:tableECDSAApproaches}
  \begin{tabular}{p{1cm} p{0.9\linewidth}}
  \toprule
Work by & Description\\ 
\midrule

2020 \cite{canetti2020uc} & \ding{50} Proposing two non-interactive threshold ECDSA schemes with the signing procedure having a (3-6)-round preprocessing phased  before receiving the message and a non-interactive signing phase in which parties generate their signature shares \ding{50} Providing proactive security for the key refreshment \ding{50} Providing the capability of identifying the party deviating from the protocol with aborting the signature generation \ding{50} Achieving UC security in ROM $\star$ Small number of rounds compared to \cite{doerner2019threshold,gennaro2018fast,lindell2018fast} $\star$ Higher communication overhead compared to \cite{gennaro2018fast,lindell2018fast} with Paillier and less compared to \cite{doerner2019threshold,lindell2018fast} with oblivious transfer \\
\midrule

2019 \cite{doerner2019threshold} & \ding{50} Proposing a ($\mathit{t, n}$) threshold ECDSA scheme with $\mathit{\log{}t+6}$ rounds of computations \ding{50} Proven security against malicious adversary with the \textit{Computational Diffie-Hellman Assumption (CDH)} assumption  \ding{50} Improving the setup model of \cite{doerner2018secure} with security against dishonest majority \ding{50} Proposing a two-party multiplication protocol based on oblivious transfer, improving the one in \cite{doerner2018secure} by 40\% through requiring parties to choose random inputs and later correcting them if necessary \ding{50} Presenting a technique to extend two-party multiplier to a t-party multiplier with $\mathit{\log{}t}$+(1 or 2) rounds \ding{50} Constructing a signing procedure consisting of three t-party multipliers responsible for converting shares from multiplicative into an additive  and producing an additive sharing from product of two additive shares $\star$ Giving benchmarks for LAN, WAN, and resource-limited devices $\star$ Achieving significant better performance for setup phase compared to \cite{gennaro2016threshold,boneh2019using}, weaker performance for ($\mathit{2, n}$) signing phase compared to \cite{doerner2018secure} and same performance for ($\mathit{2, n=24}$) signing with parallelism compared to ($2, 2$) of \cite{lindell2017fast} \\
\midrule

2018 \cite{lindell2018fast} & \ding{50} Proposing a ($\mathit{t, n}$) threshold ECDSA scheme with efficient and practical distributed key generation procedure through replacing the Paillier additively homomorphic encryption with ElGamal in-the-exponent \ding{50} Exploiting easy distribution of ElGamal keys, easy computations of elliptic curves over Paillier operations, efficient zero-knowledge in elliptic curve groups over expensive zero-knowledges in Paillier, and automatic modulo computations of all homomorphic operations in elliptic curves over the requirement of adding randomness and proving its addition with zero-knowledge in Paillier
\ding{50} Proven security against malicious adversary and static corruption with the possibility of being UC secure if the zero-knowledge proofs are UC secure $\star$ Given communication overhead comparison for different curve sizes with \cite{gennaro2016threshold} $\star$ Given signing comparison with \cite{gennaro2016threshold} for up to 20 parties 
\\
\midrule

2018 \cite{gennaro2018fast} & \ding{50} Proposing the first ($\mathit{t, n}$) threshold ECDSA for any value of $\mathit{t}$ with efficient signing and communication overhead \ding{50} Using the technique of \cite{damgaard2012implementing} and presenting a two-party protocol by which two parties having additive shares can compute pairwise multiplication of those shares \ding{50} Proven security against malicious adversary with dishonest majority $\star$ Testing the running time while excluding the network overhead, achieving significant improvement of $\mathit{29 + 24t}$ over \cite{gennaro2016threshold} with $\mathit{142 + 52t}$ and \cite{boneh2019using} with $\mathit{397 + 91t}$ milliseconds\\
\midrule

2018 \cite{doerner2018secure} & \ding{50} Proposing a new n-party key generation with five rounds for ECDSA with security proven in ROM under the \textit{Computational Diffie-Hellman (CDH)} assumption \ding{50} Proposing a ($2, 2$) ECDSA with two rounds of signing proven under the CDH assumption \ding{50} Extending the proposed ($2, 2$) scheme to ($\mathit{2, n}$) with two-round signing and five-round setup phases where all parties commit to and send a single proof of knowledge of discrete logarithm to all other parties in broadcast $\star$ Implementing the scheme using Rust $\star$ In terms of concrete signing communication cost, achieving more overhead hundreds of magnitudes in ($2, 2$) scheme compared to \cite{lindell2017fast} but achieving less overhead of 15-20 times better compared to \cite{gennaro2016threshold,boneh2019using} \\
\midrule

2017 \cite{lindell2017fast} & \ding{50} Proposing a fast and efficient two-party threshold ECDSA digital signature scheme \ding{50} Using Paillier homomorphic encryption and utilizing the fact that if one party has another one's Paillier encryption of share of the private key, the other party can participate in generating the commitment and computing a public verifiable signature with no zero-knowledge \ding{50} Having key generation more complicated than signature generation due to proving correct generation of Paillier key and bridging the gap between Paillier encryption and DLP \ding{50} Proving security for sequential composition under standard assumptions $\star$ Two times faster than previous schemes $\star$ 37ms signing time with the use of the curve P-256 on machines on the Azure infrastructure $\star$ Key generation of 2.5 seconds on machines of Azure running with a single thread \\
\midrule

2016 \cite{gennaro2016threshold} & (See \ref{sec:ecdsa} for details) \\
\bottomrule
\end{tabular}
\ktablehelp
\end{table}

\subsection{Threshold EdDSA} \label{sec:eddsa}

The EdDSA digital signature, depicted in Figure \ref{fig:SchnorrItsVariants}(b), proposed by Bernstein et al. \cite{bernstein2012high} and specified in \cite{josefsson2017edwards}, is a digital signature that utilizes Edwards curves and is based on the ECDLP (See \ref{def:ecdlp}). It is a variant of the Schnorr signature and uses performance-optimized elliptic curves specified by the NIST \cite{brandao2023notes} that includes the Edwards25519 curve for 128-bit strength and the Edwards448 curve for 224-bit strength. These elliptic curves provide two distinct signing modes, each determined by whether the message undergoes prehashing or not, allowing for four total modes. Table \ref{tab:ThreshooldEdDSA} contains a compilation of recent works on threshold EdDSA signatures.

The recent work by Battagliola et al. \cite{battagliola2020provably} proposed a multi-party EdDSA digital signature scheme provable secure against an adaptive adversary in which the involvement of all the players is not required during the key generation, and no trusted party is required in the scheme. The proposed scheme is the variant of a threshold ECDSA scheme by the same authors \cite{battagliola2022threshold} in which has been shown that it can be strengthened in the presence of a rushing adversary and can obtain resiliency of the recovery against a malicious adversary. Using a non-malleable commitment scheme, a secure hash function, and an IND-CPA encryption scheme, this paper considers a $(2,3)$ threshold scheme in which, after first initializations and communication of common parameters, one player chooses a long-term asymmetric key and goes offline. The key generation phase is played by the remaining two parties and can be played in two ways: Key generation done by the remaining two parties or done by the offline player and one of the others. The signature generation process involves the participation of two parties. To ensure determinism, Purify \cite{nick2020musig}, an EC-based pseudo-random function (PRF), is utilized to obtain verifiable nonce values. This allows each party to verify whether the other party has correctly computed the random value. Finally, when one of the two parties cannot participate in the signature generation, the offline party joins and participates in the signature recovery process.

\begin{table}[h]
\scriptsize
  \caption{Summary of Threshold EdDSA Digital Signatures}
  \label{tab:ThreshooldEdDSA}
  \begin{tabular}{p{1cm} p{0.9\linewidth}}
  \toprule

Work by & Description\\ 
\midrule

2023 \cite{feng2023efficient} & \ding{50} Proposing two practical multi-party EdDSA signature schemes for semi-honest and malicious security where the semi-honest protocol realizes the EdDSA standard and the malicious protocol provides security against a static adversary with at most $\mathit{n-1}$ corrupted parties and eliminates the distributed hashing and the MPC required for it by securely managing a global state \ding{50} Extending the malicious protocol to allow identifiable abort $\star$ Implementing the proposed protocols in C++ based on Ed25519 curve and running them with 2-5 parties on Alibaba cloud servers $\star$ 14.47 and 4.13 milliseconds for key generation and 15.3 and 10.26 milliseconds for signing in semi-honest and malicious settings with five parties, respectively\\
\midrule

2022 \cite{shi2022threshold} & \ding{50} Proposing a ($\mathit{t, n}$) EdDSA threshold signature scheme proved to be secure based on the ECDLP problem and consisting of two phases: Setup phase, in which the private key and secret vectors are secretly shared among the parties. Signing phase, in which using distributed random element generation approaches, parties can generate signatures efficiently without using ZKP and MPC  $\star$ Providing performance comparison both for the cloud and embedded implementations for the number of parties and threshold values up to 9 \\
\midrule

2020 \cite{battagliola2020provably} & Proposing a ($2, 3$) EdDSA digital signature scheme with an offline recovery party provable secure against an adaptive adversary using anon-malleable commitment scheme, a secure hash function, and an IND-CPA encryption scheme (See \ref{sec:eddsa} for details) \\
\midrule

2020 \cite{feng2020practical} & \ding{50} Proposing the first two-party EdDSA threshold signature scheme with a security reduction to the security of the original EdDSA scheme $\star$ Providing performance analysis for signature and key generations in client/server and device/device models for semi-honest and malicious security settings \\
\bottomrule
\end{tabular}
\ktablehelp
\end{table}

\vspace{-5mm}
\section{Threshold Pairing-based Digital Signatures}\label{sec:pairing}

In this section, we will briefly explore the concept of pairing-based cryptography and introduce a pairing-based signature known as BLS \cite{blsboneh2004short}. We will then describe how this signature scheme can be thresholded.

\begin{definition}[Pairings]
\label{def:pairings}
Let ($G_1$, $g_1$) and ($G_2$, $g_2$) represent two additive groups with their generators, and $G_T$ be a multiplicative group of prime order $q$. We introduce $\hat{e}$ as a \textit{bilinear map} or \textit{pairing}, denoted as $\mathit{\hat{e}: G_1 \times G_2 \rightarrow G_T}$, possessing the following properties:

\begin{enumerate}
\item[-] \textit{Bilinearity}: $\forall g_1 \in G_1, \forall g_2 \in G_2,$ and $ \forall a,b \in \mathbb{Z}^{*}_q$ $\rightarrow$ $\hat{e}(ag_1, bg_2) = \hat{e}(g_1, g_2)^{ab}$ holds.
\item[-] \textit{Non-degeneracy}: $\forall g_1 \in G_1, \forall g_2 \in G_2,$ and $g_1 \neq 0$ and $g_2 \neq 0$ $\rightarrow$ $\hat{e}(g_1, g_2)$ generates $G_T$.
\item[-] \textit{Computability}: Computation of $\hat{e}$ should be efficient.
\end{enumerate}
\end{definition}

Notable pairings in this context are the \textit{Weil pairing} \cite{menezes1991reducing} and the \textit{Tate pairing} \cite{frey1994remark}. As pairings advanced, they found applications in various cryptographic domains. For instance, pairings were utilized to create Identity-based Encryption \cite{boneh2001identity} and Public Key Encryption with Keyword Search \cite{boneh2004public}. In digital signatures, the Boneh, Lynn, and Shacham (BLS) signature scheme \cite{blsboneh2004short} was introduced and functions in any group with an easily solvable \textit{Decisional Diffi-Hellman Problem (DDH)} and a computationally hard \textit{Computational Diffie-Hellman (CDH)}. Additionally, pairings played a role in key agreement protocols, including the one-round 3-party key agreement scheme by Joux \cite{joux2000one} and its multi-party extension by Barua et al. \cite{barua2003extending}.

\begin{definition}[BLS Digital Signature]
\label{def:BLSds}
Given a computable non-degenerate pairing $\mathit{\hat{e} = G_1 \times G_2 \rightarrow G_T }$ in groups $G_1$, $G_2$, and $G_T$ of prime order q with $g_1$, $g_2$, and $g_t$ the generators of them, respectively. With $H: \mathcal{M} \rightarrow G_1$ be a Map-to-point hash function, we define the $BLS = (Setup, Kg, Sig, Ver)$ signature as:

\begin{enumerate}
\item[-] \underline{$params \leftarrow BLS.Setup(\kappa)$}: Setting up the groups $G_1$, $G_2$, and $G_T$.
\item[-] \underline{$(sk, PK) \leftarrow BLS.Kg(params)$}: Choose $sk \xleftarrow{\text{\$}} \mathbb{Z}_q $ and compute $PK \leftarrow g_2^{sk} \in G_2$.
\item[-] \underline{$\sigma \leftarrow BLS.Sig(m)$}: Given the message $m$, compute $h \leftarrow H(m) \in G_1$ and the signature $\sigma \leftarrow h^{sk}$.
\item[-] \underline{$b \leftarrow BLS.Ver(m, \sigma, PK)$}: Given the message-signature pair $(m, \sigma)$ and the public key $PK$, check if $\hat{e}(\sigma, g_2) \stackrel{?}{=} \hat{e}(H(m), PK)$ holds and output the decision bit $b$ ($b=1$ if the statements is true, otherwise 0).
\end{enumerate}
\end{definition}

A thresholded version of the BLS digital signature was first given by Boldyreva \cite{boldyreva2002threshold}. She proposed a $(t, n)$ threshold digital signature that can tolerate any $t < n/2$ malicious parties proven in ROM with the capability of achieving proactive security using \cite{herzberg1997proactive}. Moreover, the key generation process does not require a trusted dealer, and the signature generation is non-interactive and does not utilize any ZKP. Given the groups $G_1$, $G_2$, and $G_T$, the key generation is done using the DKG \cite{gennaro2007secure}, which is jointly executed by a set of parties. This results in outputting a public key $y=g^x$, and giving each party a private key share $x_i$ such that $x$ can be constructed using Shamir's secret sharing as $x \leftarrow \Sigma_{i \in R}(L_i x_i)$ where $L_i$ is Lagrange coefficient and $R$ is a set of $t+1$ players. To generate a signature, each party computes and broadcasts a signature share $\sigma_i \leftarrow H(m)^{x_i}$, independent of the number of parties. The signature can be constructed by any player(s) having shares of $R$ as $\sigma \leftarrow \Pi_{i \in R}(\sigma_i ^ {L_i})$. The verification remains the same as the BLS signature. 

Several existing literature approaches focus on thresholding or utilizing the BLS signature to create threshold systems. These include Bacho et al. \cite{bacho2022adaptive} work addressing the lack of security proof of Boldyreva's scheme \cite{boldyreva2002threshold} against adaptive security. Tomescu et al. \cite{tomescu2020towards} proposed a $\mathit{(t, n)}$ threshold BLS scheme with $\mathit{\theta(t \log{}^2 t)}$ aggregation time and constant signature size, signing, and verification times. Bellare et al. \cite{bellare2022stronger} introduced a framework for non-interactive threshold signatures, formalizing unforgeability (UF) and strong unforgeability (SUF) and demonstrating their applicability to threshold BLS and FROST. Garg et al. \cite{garg2023hints} presented a novel threshold BLS signature, ensuring proactive and forward security and featuring a silent setup.

\section{Threshold RSA Digital Signature}\label{sec:rsa}
The RSA digital signature \cite{jonsson2003public,moriarty2016pkcs,chen2023digital}, is a scheme based on the Integer Factorization problem (See \ref{def:IntFactP}) problem. While there is a compilation of works on threshold RSA signatures in Table \ref{tab:tableRSAApproaches}, the research focuses primarily on EC-based approaches due to their widespread use in applications such as cryptocurrencies. As a result, the development of threshold RSA signature schemes has received less attention in recent years. Table \ref{tab:tableRSAApproaches} contains some of the works focusing on threshold RSA signatures in addition to the works of \cite{almansa2006simplified,damgaard2005efficient}. 



\vspace{-3mm}
\begin{table}[H]
\scriptsize
  \caption{Summary of Threshold RSA Digital Signatures}
  \label{tab:tableRSAApproaches}
  \begin{tabular}{p{1cm} p{0.9\linewidth}}
  \toprule

Work by & Description\\ 
\midrule

2013 \cite{dossogne2013secure} & \ding{50} Giving a variant of the RSA threshold scheme in \cite{ghodosi2006ideal} which the simulation proofs of schemes like \cite{gennaro2008threshold,shoup2000practical} are difficult to apply since the proofs require the evaluation of non-integer coefficients which may not be able to be evaluated because the denominator is not coprime with modulus \ding{50} Provable security against static adversary and giving robust version secure against active adversary\\
\midrule

2008 \cite{gennaro2008threshold} &  \ding{50} Proposing an efficient variant of the RSA threshold signature of \cite{shoup2000practical} for a large number of parties in ad-hoc networks and making the protocol independent of the number of parties which had to be fixed and known to all parties \ding{50} Extending the proposed threshold scheme to support dynamic groups without trusted center using the non-interactive solution \cite{saxena2005efficient} \ding{50} Proving security against a passive adversary in ROM and extending robustness to tolerate malicious adversary \\
\midrule

2000 \cite{shoup2000practical}  & \ding{50} Proposing a new robust non-interactive RSA threshold signature with provable security against a static adversary in ROM that size of signature share is a multiplication of a constant and RSA modulus \ding{50} Requiring a trusted center since the RSA modulus is a product of two safe primes \ding{50} Generalizing the concept of thresholding as $\mathit{t}$ denotes the number of corrupted parties and $\mathit{k}$ denotes the number of shares required for signature generation and proving the security with $\mathit{k}=\mathit{t}+1$ \ding{50} Proposing a more general protocol to be used when $\mathit{k} \ge \mathit{t}+1$ proven in ROM \\

\bottomrule
\end{tabular}
\ktablehelp
\end{table}

\vspace{-3mm}
\begin{table}[H]
\scriptsize
  \caption{Summary of Threshold RSA Digital Signatures (Cont.)}
  \label{tab:tableRSAApproaches}
  \begin{tabular}{p{1cm} p{0.9\linewidth}}
  \toprule

Work by & Description\\ 
\midrule

1998 \cite{rabin1998simplified} & \ding{50} Proposing a general paradigm for transforming an all-player protocol into a threshold one \ding{50} Utilizing the proposed paradigm with RSA signature scheme to obtain interactive threshold RSA \ding{50} Distributing the RSA secret key as \textit{additive-shares} \ding{50} Adding robustness by further sharing the share secretly using a secure VSS \ding{50} Generating signature by multiplying the partial signatures of the parties with possibility of finding malicious party upon facing invalid signature \ding{50} Achieving proactive RSA by changing the representation of the secret key   \\
\bottomrule
\end{tabular}
\ktablehelp
\end{table}

%% file: bodies/generic_approaches.tex
The properties and security models of MPC were discussed in Section \ref{sec:mpcdef}. MPC's inherent functionality makes it well-suited for threshold cryptography, particularly in thresholding digital signatures. By employing MPC-based methods, we can achieve \textit{Transparent thresholding}, which allows converting existing signatures into thresholded schemes while preserving their underlying properties. Using MPC offers the advantage of removing dependencies on the underlying construction of the scheme. However, it also brings disadvantages like higher communication overhead and computational costs, making implementing it in resource-limited applications challenging. In the following, we will explore the application of MPC in thresholding conventional signatures. Further, in Section \ref{sec:PQC}, we will investigate the application of MPC in thresholding the NIST's standardized post-quantum digital signatures.

\subsection{Thresholded MPC Protocols}
MPC encompasses a range of protocols, and in most modern actively-secure protocols, the generation of random data or utilizing private information is often required to be secretly shared. Traditional methods rely on \textit{Verifiable Secret Sharing (VSS)} in an information-theoretic model, with a corruption threshold of less than $\frac{n}{3}$ parties. However, more recent protocols operate under a \textit{Full threshold} setting, as incorporated by the SPDZ family \cite{damgaard2012multiparty}, Tiny-OT family \cite{larraia2014dishonest,nielsen2012new}, SPDZ-2k \cite{cramer2018spd}, and "n-party GC" family \cite{hazay2020low,wang2017global} where up to $n-1$ parties can be corrupted. Specific protocols like "Special GC" \cite{mohassel2015fast} achieve a threshold of $(t=1, n=3)$, while "General Q2" \cite{smart2019error} employs a $Q_2$ access structure with $t < \frac{n}{2}$ to ensure that no combination of two unqualified subsets can reconstruct the complete set.

\subsection{Transparent Thresholding of Conventional Digital Signatures}
In the upcoming subsections, we will explore the transparent thresholding of some standard digital signatures using MPC. Table \ref{tab:mpcBasedTDS} provides an overview of these schemes.

\subsubsection{MPC Schnorr}
The Schnorr signature and its derivatives (e.g., ECDSA and EdDSA) use confidential elements that support linear combinations. This property enables the creation of multi-party signature variants by applying linear secret sharing methods to these elements. Generating a fresh random value (nonce) for each signature in the original Schnorr signature presents a challenge. Normally, this value is chosen randomly from a specific set ($\kcyczoverq$). However, there is no guarantee of a consistently reliable source of entropy, unlike EdDSA, which deterministically calculates the nonce based on the secret key and the message. One solution to tackle the entropy problem is to employ \textit{State continuity}. This involves designing a randomized protocol and utilizing a block cipher with a fresh counter to derive a new random value. However, Parno et al. \cite{parno2011memoir} have demonstrated that even highly isolated devices cannot be fully trusted to maintain the state accurately due to software errors, power interruptions, or malicious attacks.

Garillot et al. \cite{garillot2021threshold} proposed a threshold Schnorr signature that employs a stateless nonce derivation method by first introducing a zero-knowledge proof system based on the \textit{Zero-knowledge from the Garbled Circuits (ZKGC)} paradigm of Jawurek et al. \cite{jawurek2013zero}. This system eliminates the need for a long-term state and maintains efficiency with at most three AES invocations per AND gate in the circuit. However, two challenges arise. The first challenge is efficiently handling a large number of oblivious transfers ($O(q)$) and public key operations per invocation required for an input of size $q$ bits. This challenge has been addressed by a variant of \textit{Committed Oblivious Transfer} enabling the prover to commit to a witness once but prove multiple statements. This technique is almost 11.5 times faster than using standard OT, but its communication overhead is five times more. The second challenge is the computational cost of relating elliptic curve group statements to a boolean circuit using a method from Chase et al. \cite{chase2016efficient}. This challenge was mitigated by a gadget for garbling the exponentiation function that takes an encoded string in Yao's garbled circuits style as input and outputs a convenient algebraic encoding of that value. This gadget is 128 times cheaper than the approach proposed in \cite{chase2016efficient} in terms of communication overhead and requires 62 times fewer calls to the cipher used for garbling. The threshold scheme involves parties sampling initial states during key generation to deterministically derive signing confidential information upon receiving a message. This determinism is achieved by using a PRF, sampled in the first step, applied to the message. However, this approach is not secure against active adversaries. To address this security concern, parties must commit to initial random values and provide proof of the correct computation of each message relative to the committed randomness, utilizing the GMW protocol. This ensures the protocol achieves malicious security without requiring an honest majority security model. By employing this method, the Schnorr signature can be transformed into a thresholded version, where parties commit to a PRF and provide proof that the discrete logarithm of their claimed nonce has been deterministically computed using that PRF.

\subsubsection{MPC ECDSA}

Dalskov et al. \cite{dalskov2020securing} first proposed a generic transformation from MPC protocols over a field $\mathbb{Z}_p$ to protocols over an elliptic curve of order $p$. This was done via introducing a new \textit{Arithmetic black-box functionality (ABB)} functionality where the commands defined in this functionality are assumed to be supported by the MPC. Further, this ABB was extended to support secure computation over an arbitrary abelian group of order $p$ to perform MPC in subgroup $G$. Extended ABB is sufficient to give a protocol over a group, properties through which linearity is preserved, secret reconstruction is relied only on the group operations in $\mathbb{Z}_p$, and multiplication of secret with public point is possible. Then, they used MP-SPDZ \cite{keller2020mp} to enable threshold ECDSA signing within different threat models by showing that the MAC scheme of SPDZ can be used to provide share authentication.

\subsubsection{MPC EdDSA}

MPC facilitates bit-wise operations on secret values within a finite field or ring. However, when applied to shared secrets (SS-based MPC), it requires complex preprocessing and increases communication rounds. Recent research by \cite{keller2018efficient} has demonstrated multi-party boolean circuit garbling with active security in the dishonest majority. Nevertheless, these methods can be computationally expensive due to modulo reduction in the circuit or relying on inefficient two-party protocols for achieving even passive security. To tackle these challenges, mixed protocols have emerged. These protocols enable parties to switch between SS-based MPC, which offers efficiency for basic addition and multiplication operations, and GC-based MPC, which provides lower round complexity. This transformation between a SS share and a GC share can be done using \textit{Double-shared authenticated bits (daBit)} \cite{rotaru2019marbled}.

The use of hash functions in certain schemes poses a challenge when attempting to create threshold versions of these schemes, as it can undermine the crucial property of linearity in the secrets. The EdDSA scheme serves as an example, employing hash functions on the secret key and the message to generate an ephemeral secret key. There are two variants of EdDSA based on the application of the hash function. The first variant involves hashing the secret key and rehashing it with the plain message to calculate the ephemeral key as $r = H(H(sk) || m)$. The second variant, called HashEdDSA, applies the hash function to both the key and the message, resulting in $r = H(H(sk) || H(m))$. The latter approach is more suitable for thresholding since computing hash functions using MPC incurs significant computational costs. Consequently, utilizing a shorter input reduces the complexity of the threshold computation.

Bonte et al. \cite{bonte2021thresholdizing} demonstrated that it is feasible to achieve a threshold version of a specific variant of EdDSA by making slight modifications to existing MPC techniques. They leverage the work presented in \cite{smart2019distributing} to employ MPC for elliptic-curve-based computations on the field $\mathbb{F}_q$, where the ephemeral key is randomly selected. For computing the hash functions, they followed two approaches. The first approach is a GC-based method, utilizing a variant of the HSS protocol \cite{hazay2020low}, which enables the generation of additive authenticated bit-wise sharing of the hash outputs among the parties. The second approach is an SS-based method, inspired by the work of Araki et al. \cite{araki2017optimized}, which operates over the finite field $\mathbb{F}_2$ which further by employing a daBit, they converted the bit-sharings into an $\mathbb{F}_q$-sharing format suitable for the $Q_2$ access structure. This access structure offers advantages over full threshold access structures by reducing the complexity of MPC and allowing to locally compute the additive sharing of a product of two secrets.

\begin{table}[H]
\scriptsize
  \caption{Summary of MPC-based Threshold Digital Signatures}
  \label{tab:mpcBasedTDS}
  \begin{tabular}{ p{1cm} p{0.1\linewidth} p{0.19\linewidth} p{0.5\linewidth}}
    \toprule
Work by & Scheme & Threshold & Security\\ 
\midrule
2022 \cite {lindell2022simple} & Schnorr, EdDSA & $\mathit{(t, n)}$ & Standard real/ideal paradigm, Proactive adversarial model, UC-security, Identifiable abort\\
2022 \cite {meier2022mpc} & Schnorr & $\mathit{(t, n)}$ & MPC with commitment, UC-security, Identifiable abort\\
2021 \cite{garillot2021threshold} & Schnorr & $\mathit{(t, n)}$ & Dishonest majority, UC Commitments, Zero-knowledge from Garbled Circuits (ZKGC)\\
2021 \cite{bonte2021thresholdizing} & EdDSA & $\mathit{Q_2}$ access structure ($t<n/2$) & Active adversarial model \\
2020 \cite{dalskov2020securing}  &  ECDSA  & $\mathit{(t,n=3)}$ & Security of the underlying ABB and assumption of ECDSA to be secure   \\ 
\bottomrule
\end{tabular}
\end{table}

%% file: bodies/PQC.tex
In 2017, NIST \cite{nistCallProposals} initiated efforts to standardize secure digital signatures that are resistant to quantum attacks. After three rounds of evaluation \cite{nistSelectedAlgorithms}, four schemes were selected for standardization. CRYSTALS-KYBER \cite{avanzi2017crystals} was chosen for public-key encryption and key establishment, while CRYSTALS-Dilithium \cite{ducas2018crystals}, FALCON \cite{fouque2019fast}, and SPHINCS+ \cite{aumasson2019sphincs} were selected for digital signatures. Moreover, other organizations have also made notable contributions to the advancement of post-quantum cryptography, such as the \textit{Internet Engineering Task  Force (IETF)} focusing on Transport Layer Security (TLS) and Internet Key Exchange (IKE), the \textit{European Telecommunications Standards Institute (ETSI)} concentrating on Quantum Key Distribution (QKD)~\cite{Yavuz:22:QKD} and PQC with the key exchange standard \cite{ETSICYBER}, and the \textit{International Standards Organization (ISO)} discussing post-quantum algorithms in SD8 of JTC 1/SC 27. China, alongside the United States and Europe, has actively engaged in the standardization of post-quantum schemes. The LAC scheme \cite{lu2018lac}, accepted in a competition organized by the \textit{Chinese Association for Cryptographic Research}, reached the second round of the NIST's standardization. To enable thresholding in PQ-secure digital signatures, a combination of MPC techniques like \textit{Linear Shamir's secret sharing (LSSS)}-based MPC and \textit{Garbled circuit (GC)}-based MPC is required.

After announcing the candidates for standardization, NIST declared the continuation of the PQC standardization process with a fourth round \cite{nistStandardizationAdditional}. To preserve diversity in standardized signatures, NIST primarily solicits additional post-quantum secure digital signatures based on structures other than lattices. This call aims to benefit applications requiring short signatures and fast verification processes. While the submission of lattice-based signatures, such as HAETAE \cite{haetae} and Raccoon \cite{raccoon} inspired by Dilithium, and SQUIRRELS \cite{squirrel} inspired by FALCON, is permitted, such submissions are required to outperform previous schemes or include additional security features.

In the following sections, our focus will primarily be on the lattice-based and hash-based approaches employed in the standardized schemes. This study does not cover other approaches, such as isogeny or code-based approaches, which are the basis for non-standardized schemes in previous evaluation rounds.

\subsection{Threshold Lattice-based Digital Signatures}
Lattice-based digital signatures have been the focus of extensive research since their introduction by Ajtai \cite{ajtai1996generating}. However, many proposed schemes suffer from security vulnerabilities due to private key leakage through signatures. To overcome this, Lyubashevsky et al. \cite{lyubashevsky2009fiat} introduced the \textit{Fiat-Shamir-with-aborts} methodology to address key leakage in lattice-based schemes. Moreover, rejection sampling is a significant challenge in lattice-based schemes, where intermediate values must be kept confidential until sampling is complete. This requires both GC-based MPC for non-linear operations and LSSS-based MPC for linear operations, along with the required transformation using daBits. These factors pose challenges in achieving thresholding in lattice-based schemes. In the following, we will investigate the thresholding of the NIST's lattice-based digital signature candidates, while Table \ref{tab:tableLatticePQschemes} provides a list of additional non-standard lattice-based digital signatures that have been thresholded.

\begin{figure}[H]
\includegraphics[width=155mm, scale=2]{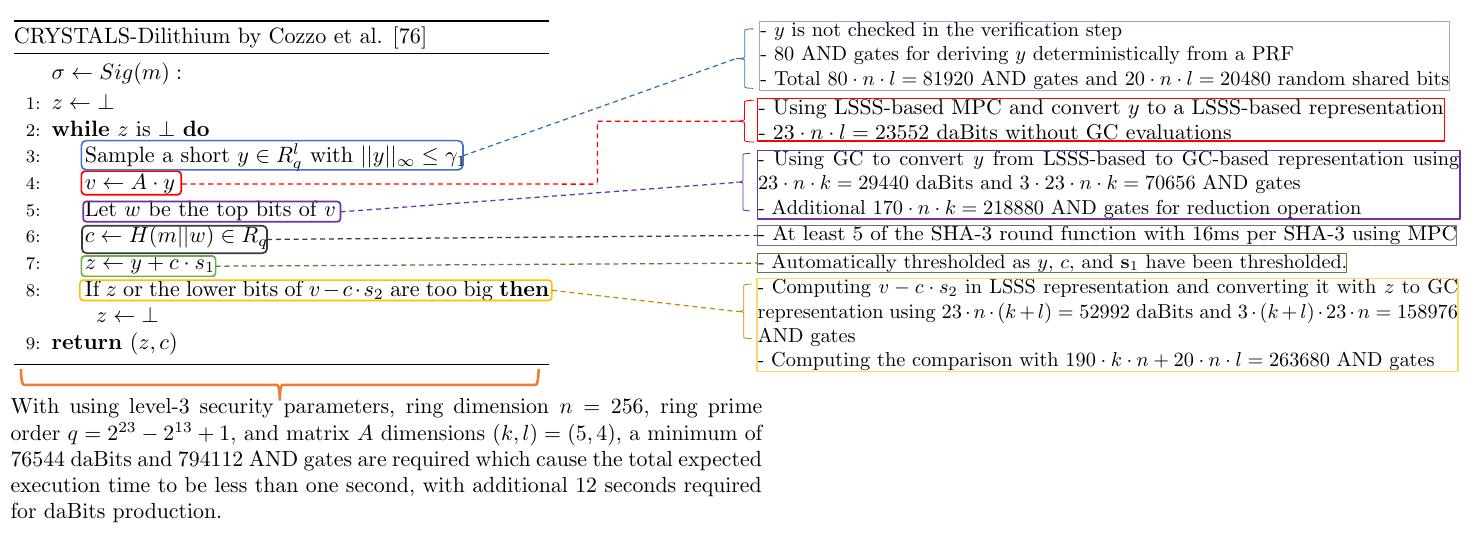}
\caption{Threshold CRYSTALS-Dilithium Digital Signature (Level-3 security parameters)}
\label{algo:DilithiumColor}
\end{figure}

\paragraph{\textit{CRYSTALS-Dilithium}} CRYSTALS-Dilithium \cite{ducas2018crystals} is a PQ digital signature that incorporates \textit{Multiplication in Polynomial Rings} and \textit{XOF (Extendable-output Function) Expansion} as building blocks. This signature is based on the M-LWE problem (See \ref{def:mlwe}) and follows the Fiat-Shamir-with-aborts paradigm, allowing for convenient adjustment of the security level. Unlike previous lattice-based schemes \cite{ducas2013lattice,ducas2014efficient} that used \textit{Gaussian sampling} and were vulnerable to side-channel attacks, this scheme employs uniform sampling with constant-time operations to improve its security and reduce the risk of side-channel attacks. Additionally, this scheme achieves a more efficient design by avoiding Gaussian sampling, reducing overhead for transmitting public keys and signatures. A simplified depiction of the protocol proposed by Cozzo et al. \cite{cozzo2019sharing} can be found in Figure \ref{algo:DilithiumColor}, with the cost of thresholding of each step and the entire scheme.


\begin{figure}[H]
\includegraphics[width=155mm, scale=2]{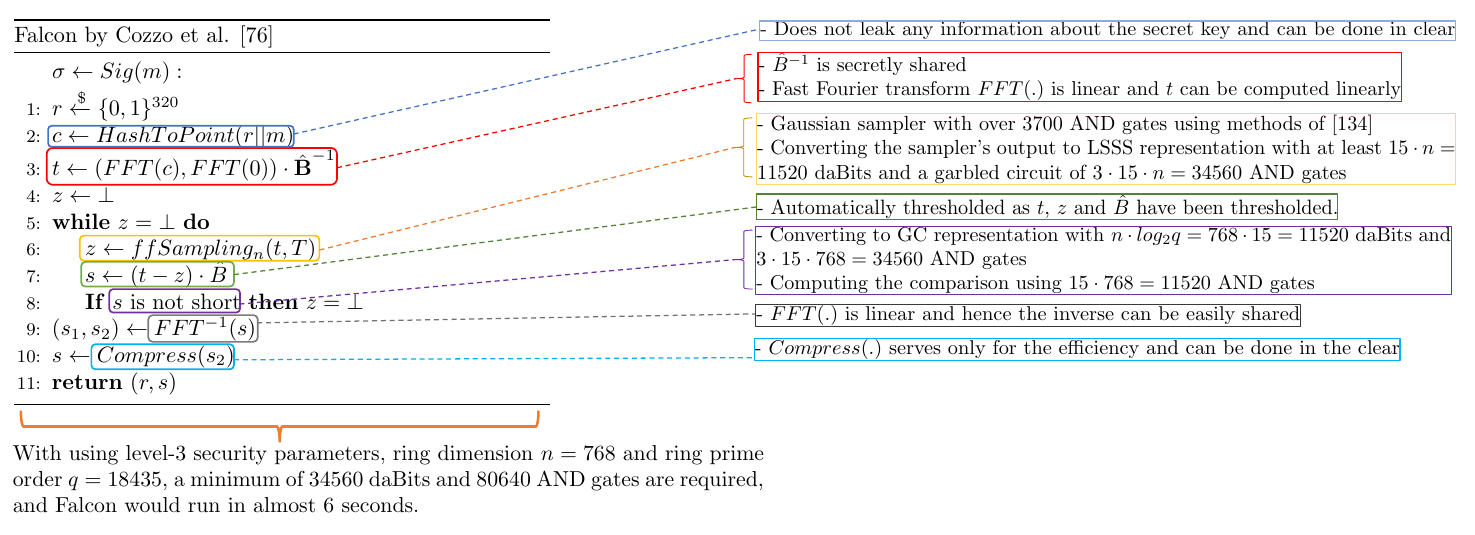}
\caption{Threshold FALCON Digital Signature (Level-3 security parameters)}
\label{algo:Falconcolor}
\end{figure}

\paragraph{\textit{{FALCON}}} The FALCON signature \cite{fouque2018falcon} is a compact PQ digital signature. Unlike CRYSTALS-Dilithium, FALCON adopts the \textit{hash-and-sign} paradigm over Fiat-Shamir to benefit from the GPV framework \cite{gentry2008trapdoors}, which provides security in both classical and quantum oracle models, as well as message recovery capability \cite{del2016whole}. NTRU lattices within the GPV framework allow for compact instantiation and accelerated operations due to their ring structure. A simplified depiction of the protocol proposed by Cozzo et al. \cite{cozzo2019sharing} can be found in Figure \ref{algo:Falconcolor}(left), along with the cost of thresholding of each step and the entire scheme. The algorithm utilizes the $FFT$ procedure to compute the Fast Fourier Transform of a given polynomial. It employs matrix $\textbf{B}$ to store randomly generated polynomials during key generation. The $ffSampling$ applies randomized rounding on the coefficients of $t$ and the given FALCON tree $T$.

\vspace{-3mm}
\begin{table}[h]
\scriptsize
  \caption{Summary of Threshold Lattice-based Digital Signatures}
  \label{tab:tableLatticePQschemes}
  \begin{tabular}{p{1cm} p{0.9\linewidth}}
  \toprule

Work by & Description\\ 
\midrule

2022 \cite{agrawal2022round} & \ding{50} Optimal improving the existing threshold scheme proposed by \cite{boneh2018threshold} in various ways: 1. Reducing noise flooding ratio from exponential to sub-linear complexity and hence, increasing the efficiency 2. Addressing the gap in clarification of the underlying signature scheme using a variant of Lyubashevsky's signature \cite{lyubashevsky2012lattice} that does not rely on rejection sampling 3. Achieving \textit{Partial adaptivity} in the ROM and \textit{Full adaptivity} in the standard model with having preprocessing \ding{50} Giving a general ($\mathit{t, n}$) access structure using FHE, UF-CMA signature scheme, and \{0, 1\}-LSSS \\
\midrule

2021 \cite{pilaram2021efficient} & \ding{50} Improving their previous lattice-based threshold multi‐stage secret sharing (TMSSS) scheme \cite{pilaram2015efficient} such that less public values are published \ding{50} Proposing a new anonymous and verifiable threshold signature scheme based on the trapdoor function of the \cite{micciancio2012trapdoors} and the improved TMSSS where a combiner, leveraging additive homomorphic property of TMSSS, combines partial signatures of the parties and generates a signature $\star$ Improving the share to secret size ratio of the best existing scheme by a factor of two \\
\midrule 

2021 \cite{vakarjuk2021dilizium}  & \ding{50} Proposing a three-round lattice-based two-party ($2, 2$) signature scheme constructed from a (bit decomposition)-free variant of the Crystals-Dilithium post-quantum signature scheme \cite{kiltz2018concrete} that follows Fiat–Shamir with Abort paradigm with security based on M-LWE and M-SIS (See \ref{def:mlwe} and \ref{def:msis}) proven in the random oracle model \ding{50} Using uniform distribution instead of discrete Gaussian distribution due to ease of secure implementation \ding{50} Using homomorphic hash function instead of homomorphic commitments for efficiency $\star$ Significant improvement in the size of the data communicated in the rounds $\star$ No signing comparison due to lack of implementation of other schemes \\
\midrule

2018 \cite{boneh2018threshold} & \ding{50} Constructing a threshold FHE (TFHE) with a trusted party based on LWE using two methods of LSSS and plain SSS while managing the noise blow-up \ding{50} Constructing a three-procedure universal thresholdizer based on the proposed TFHE for thresholding cryptographic primitives including public-key encryption (PKE), PRFs, and signatures schemes allowing single-round threshold signature and single-round threshold CCA-secure PKE schemes \ding{50} Constructing a five-procedure threshold signature and proving the unforgeability, robustness, and compactness \ding{50} Proving security using different hybrid games  $\star$ Compact ciphertext with key share size overhead for \{0, 1\}-LSSS $\star$ Increasing modulus overhead with more compact key shares for SSS\\
\midrule

2015 \cite{wang2015new} & \ding{50} Constructing a lattice-based threshold attribute-based (TABE) signature scheme based on SIS \ding{50} Incorporating a signature constructed from a proof of knowledge protocol using \textit{Fiat-Shamir heuristic} \cite{fiat1986prove} \ding{50} Proven selective-predicate and adaptive-message unforgeability in ROM  $\star$ Improved private key storage cost with comparable public key and signature sizes \\
\midrule

2014 \cite{choi2014lattice} & \ding{50} Relying on \textit{hash-and-sign} signature scheme and proposing \textit{Message Block Sharing (MBS)} as a tool to construct a ($\mathit{t, n}$) threshold signature scheme by which a large message is divided into randomized size portions with further giving to the parties using a combinatorial way \ding{50} Constructing a ($2, 3$) MBS and further generalizing it to ($\mathit{t, n}$) \ding{50} Incorporating the MSB into the lattice-based group signature scheme of \cite{gordon2010group} to build a threshold signature scheme $\star$ Possible inefficiency due to generating group size after receiving message for signing \\


\bottomrule
\end{tabular}
\ktablehelp
\end{table}

\vspace{-4mm}
\subsection{Threshold Hash-based Digital Signatures}
SPHINCS+ proposed by Aumasson et al. \cite{aumasson2019sphincs} is a stateless hash-based post-quantum digital signature that builds upon SPHINCS \cite{bernstein2015sphincs} scheme which incorporates various components of XMSS \cite{buchmann2011xmss}, the first post-quantum signature scheme. SPHINCS+ authenticates significant number \textit{Few-time Signatures (FTS)} using a hypertree structure composed of \textit{Many-time Signatures (MTS)}, each consisting of a Merkle-tree signature encompassing a \textit{One-time Signatures (OTS)} with an authentication path. In terms of keys, the public key is a single hash value derived from the top-level MTS, while the private key is a secret seed value used to generate all FTS and OTS keys.

Based on the performance analysis of SPHINCS+ given by \cite{cozzo2019sharing}, the total number of SHA-3 calls in SPHINCS+ is 321553. With efficient garbled circuit implementations, the execution time for these calls in the thresholded scheme is estimated to be around 85 minutes. However, following the given order of operations could significantly prolong the execution time by several hundred minutes.

%% file: bodies/special_schemes.tex
In traditional digital signatures, a single signer signs a message, and one or more recipients verify the signature. In threshold digital signatures, private data (e.g., the private key) is secretly shared among the signers, and a subset of the parties collaborates to generate the signature while the verification process remains the same. In the following, we will explore other distributed signatures, specifically Group, Ring, and Multi-signatures.

\subsection{Group Signatures} \label{sec:groupSig}
Group signatures involve a group of parties, each possessing their own signing key, who can generate digital signatures on behalf of the group while maintaining a certain level of \textit{Anonymity} and \textit{Traceability}. The signatures are verifiable using the group's single public key by any group member. If the group members are fixed, the group signature is \textit{Static}, and if it allows a user to join and leave, it is \textit{Dynamic}. Setting up and management of the group are all managed by a trusted party known as \textit{Group manager}.

The concept of group signatures was initially proposed by Chaum and Van Heyst \cite{chaum1991group}, where they introduced four group signatures based on Integer Factorization and DLP. Two of these schemes required pre-fixed group members, while the other two allowed members to select a subset from an existing fixed group to form a new group. Camenisch \cite{camenisch1997efficient} introduced dynamic group signatures that featured linear-sized group public keys, signatures, and group operations. 
Since introducing the first schemes, many other schemes have been constructed to improve the previous ones. Ateniese et al. \cite{ateniese2000practical} were the first to introduce essential properties of \textit{Collusion resistance}, \textit{Security against framing attacks}, and \textit{Exculpability} for group signatures. Bresson et al. \cite{bresson2001efficient} proposed the first revocation method with multi-revocation capability. Bellare et al. \cite{bellare2003foundations} provided a formal definition of security notions and foundations for group signatures and introduced properties such as \textit{Full anonymity} and \textit{Full traceability}. Boneh et al. \cite{boneh2004group} introduced Verifier-local Revocation (VLR) and proposed a short group signature based on this notion in which signers were not directly engaged in the revocation process, and revocation messages are exclusively sent to verifiers. Kiayias et al. \cite{kiayias2005group} presented the first group signature that utilized the BB digital signature \cite{boneh2004short} and verifiable encryption based on Paillier encryption \cite{camenisch2003practical} that allowed for concurrent joining of members using the \textit{single message and signature response} paradigm. Delerabl et al. \cite{delerablee2006dynamic} improved upon the previous works by addressing the long signature size of \cite{kiayias2005group}, the computational cost of \cite{nguyen2004efficient}, and the absence of a revocation procedure in \cite{bellare2003foundations} model. They developed a group signature that achieved signatures 70\% shorter than previous efficient schemes \cite{nguyen2004efficient}. Groth et al. \cite{groth2007fully} developed a dynamic group signature with constant size group's public key and signature using certified signatures \cite{boldyreva2007closer}.

With all the foundations of group signatures and improvements discussed earlier, certain studies have concentrated on constructing and improving threshold group signatures \cite{camenisch2020short,guo2022cryptanalysis,boneh2022threshold}. Moreover, recent works are mainly trying to utilize lattices to achieve post-quantum resistant group signatures, such as works \cite{canard2020constant,cao2022forward} to obtain static group signatures, works \cite{zhang2021improved,zhang2022verifier} to create group signatures with VLR, works \cite{kansal2020group,csahin2022constant} to achieve partially-dynamic and works \cite{kansal2020group,sun2021efficient} to achieve fully-dynamic group signatures.


\subsection{Ring Signatures} \label{sec:ringSig}
Ring signatures enable the creation of digital signatures without setting up procedures, group managers, or revocation mechanisms and offer signer anonymity within a defined group of potential signers. In these schemes, each participant possesses a private and public key. When signing a message, each party selects a subset of public keys, including their own, to form an anonymous group known as a \textit{ring}. This ring is transmitted alongside the signature. During verification, the recipient uses the ring to establish that one of the ring members generated the signature.

Rivest et al. \cite{rivest2001leak} introduced the initial formulation of a ring signature, which relied on the RSA cryptosystem. On the other hand, Abe et al. \cite{abe20021} constructed the first scheme based on the DLP. Naor \cite{naor2002deniable} developed the first interactive undeniable ring signature by combining a secure encryption scheme \cite{rivest2001leak} and the notion of \textit{Deniable Authentication} \cite{dwork2004concurrent}. Naor further extended the scheme to support the threshold ($k, n$) scheme. Zhang et al. \cite{zhang2002id} proposed the first identity-based ring signature that utilized bilinear pairings. The scheme could operate on either supersingular elliptic curves or hyperelliptic curves and could be computed efficiently using techniques described by \cite{boneh2001identity,galbraith2002implementing}. In 2003, Boneh et al. \cite{boneh2003aggregate} presented aggregate signatures, which were constructed based on \textit{Gap groups} and further utilized this concept to construct ring signatures. Efficient authentication techniques using aggregate signatures can be found in \cite{yavuz2017real,ozmen2019fast}. Dodis et al. \cite{dodis2004anonymous} presented the \textit{Ad Hoc Anonymous Identification (AHAI)} primitive, allowing users to form ad hoc groups without relying on a trusted manager while preserving member anonymity. They further developed a ring signature scheme using the AHAI primitive, which achieved a constant signature size and constant time signature generation and verification. 
Bender et al. \cite{bender2006ring} introduced the first ring signature in the standard model, inspired by the group signature of Bellare et al. \cite{bellare2003foundations}. Their work addressed the shortcomings of existing security definitions and presented a hierarchical framework of security definitions to better assess these properties in ring signatures.

Recent studies have concentrated on constructing threshold ring signatures \cite{okamoto2018k,munch2021stronger,avitabile2023extendable}. Moreover, recent advancements in ring signatures have focused on developing post-quantum resistant schemes. Esgin et al. \cite{esgin2019short} introduced the first practical lattice-based ring signature and further enhanced it in subsequent work \cite{esgin2022matrict}. Works \cite{beullens2020calamari,lyubashevsky2021smile} offered improved scalability in terms of signature size and provided concrete signature sizes only for large ring sizes, unlike \cite{esgin2019matrictMatRiCT}, which maintains efficiency across different ring sizes. Works \cite{yuen2021dualring} presented linear-size ring signatures. Katz et al. \cite{katz2018improved} proposed a hash-based approach that utilizes non-interactive zero-knowledge proofs of knowledge (NIZKPoKs) to construct a ring signature.

\subsection{Multi-Signatures} \label{sec:multiSig}
Multi-signatures involve a subset of players collaborating to create a signature, while each member in the subgroup has their own private and public key pair. These schemes allow the verifier to confirm the participation of all subgroup members. Unlike threshold digital signatures that require a predefined subgroup size, multi-signatures have no limitations on subgroup size. Unlike group signatures and ring signatures, where a single signer can generate a signature on behalf of the entire group or ring, multi-signatures require the participation of all subgroup members.

The initial concept of multi-signatures was introduced by Itakura et al. \cite{itakura1983public}. Boldyreva \cite{boldyreva2002threshold} proposed a multi-signature that relied on BLS where the requirement of prior subgroup composition before computing signature shares was eliminated. Additionally, the length of the signature and the verification process became independent of the subgroup size, resembling the properties of BLS signatures. In contrast to the approach by Micali et al. \cite{micali2001accountable}, Boldyreva's scheme allowed signers to start a new signing protocol even if a previous one was incomplete. This was due to not utilizing rewinding, which is incompatible with concurrent execution. In 2019, Maxwell et al. \cite{maxwell2019simple} introduced MuSig, a three-round Schnorr-based multi-signature that improves upon existing works and enables key aggregation. MuSig is the first multi-signature that guarantees security under the discrete logarithm assumption in the plain model. Nick et al. \cite{nick2020musig} addressed the security concerns related to the use of derandomization techniques \cite{pornin2013deterministic} and proposed a modified version called MuSig-DN, which was the first Schnorr multi-signature with deterministic signing, where the signers generate their nonce deterministically using a PRF over the message and other signers' public keys. Further, they presented the first Schnorr multi-signature, MuSig2 \cite{nick2021musig2}, that integrates key aggregation with two-round functionality, security under concurrent signing sessions, and retaining the signing complexity of the original Schnorr.

Recent research has been dedicated to developing post-quantum resistant multi-signatures. Major works have been on lattice-based approaches based on the M-LWE and M-SIS (See \ref{def:mlwe} and \ref{def:msis}) problems. Work \cite{el2016efficient} introduced a lattice-based three-round multi-signature that uses GLP signature \cite{guneysu2012practical} and is based on the Ring-LWE \cite{lyubashevsky2010ideal} and Ring-SIS \cite{lyubashevsky2010ideal}. Work \cite{damgaard2022two} addressed security concerns in \cite{el2016efficient} and proposed two alternative multi-signatures. However, these schemes exhibited reduced efficiency and longer signature lengths compared to work \cite{ma2019practical} that achieved short signatures. Moreover, the work \cite{boschini2022musig} presented a scheme that, unlike \cite{damgaard2022two}, does not use lattice-based trapdoor commitments.


%% file: bodies/applications.tex
\begin{figure}[H]
\includegraphics[width=\textwidth , scale=0.35]{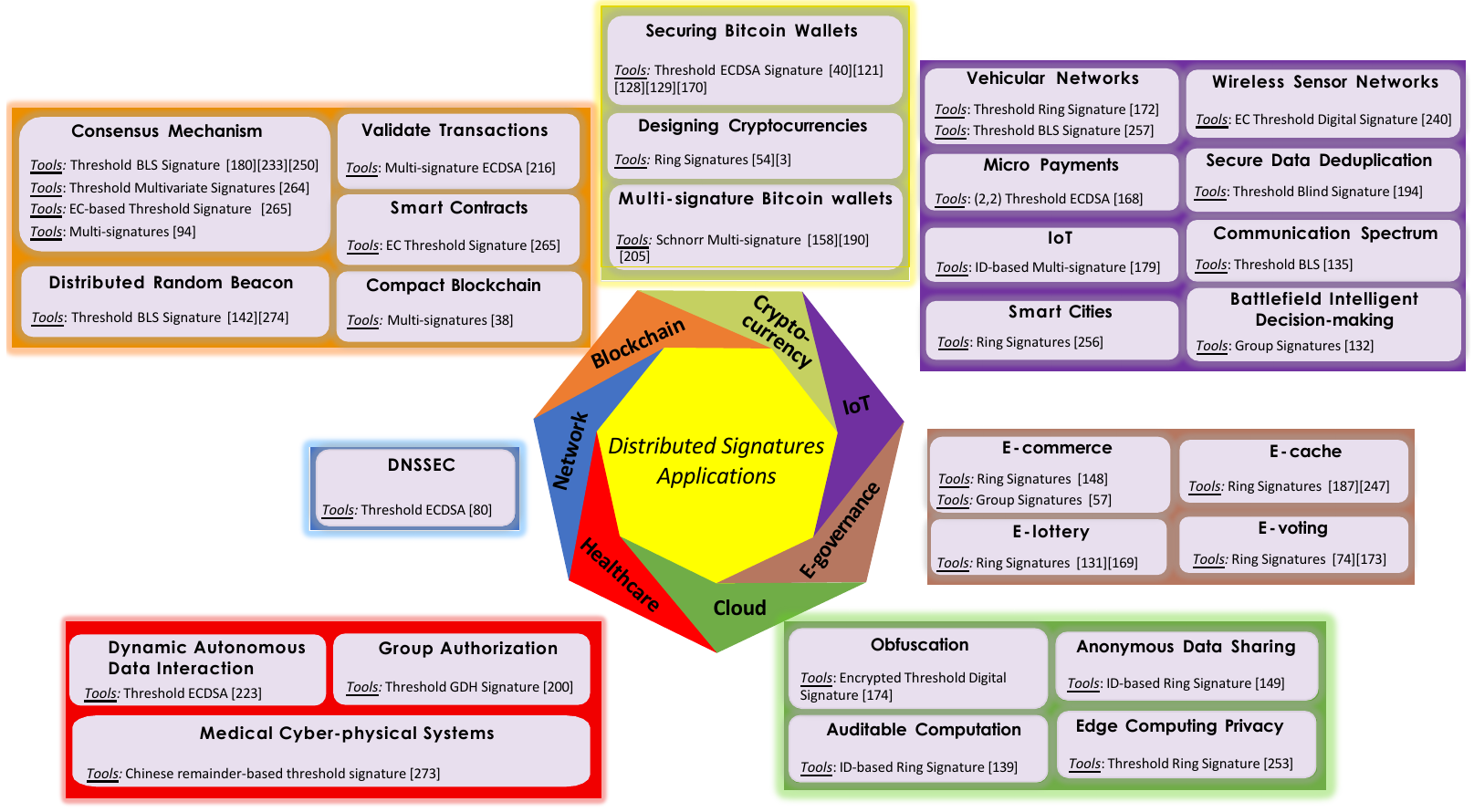}
\caption{Applications of Distributed Digital Signatures}
\label{fig:apptaxonomy}
\end{figure}

Distributed digital signatures, including threshold and exotic signatures, have practical uses in various real-world contexts. 
These applications span multiple domains, such as Blockchain, Cryptocurrency, Healthcare, IoT, etc., and have been illustrated in the taxonomy Figure \ref{fig:apptaxonomy}. In the following, we will discuss some of these applications.
                      
\underline{Blockchain:}
Effectively utilizing distributed signatures enables establishing a decentralized consensus mechanisms. These solutions encompass using multi-signatures \cite{drijvers2020pixel}, threshold BLS \cite{santiago2021concordia,liu2020sshc,wang2019smchain}, along with other threshold signatures \cite{yi2021efficient,yu2022elliptic}. Ensuring the presence of distributed randomness using \textit{Distributed random beacons} can be facilitated using threshold BLS \cite{zhu2022distributed,hanke2018dfinity}. Additionally, transaction validation can be accomplished through ECDSA multi-signature \cite{pan2022multi}, and the construction of \textit{Compact Blockchains} can benefit from the utilization of multi-signatures \cite{boneh2018compact}. Moreover, a potential approach in smart contracts involves employing threshold EC-based signatures \cite{yu2022elliptic}.

\underline{Cryptocurrency:}
Significant attention has been given to the threshold ECDSA as the primary signature scheme used in cryptocurrencies such as Bitcoin \cite{nakamoto2008bitcoin}. Works \cite{kzen2021bitcoin,boneh2019using,goldfeder2014securing,goldfeder2015securing,gennaro2016threshold} are the efforts for thresholding ECDSA, primarily centered on securing Bitcoin wallets. 
Multi-signature wallets can also be constructed and can benefit from Schnorr-based multi-signatures \cite{maxwell2019simple,nick2021musig2,kara2023efficient}. Moreover, exotic digital signatures such as ring signatures can be used for designing cryptocurrencies such as Bytecoin \cite{bytecoinWhatBytecoin} and Aeon \cite{aeonAEONPrivate}.

\underline{IoT:} 
Linkable ring signatures have been notably employed in vehicular networks \cite{li2018creditcoin} and smart cities \cite{yadav2021linkable}. For facilitating threshold computations in micro payments, a threshold $(2, 2)$ ECDSA can be utilized \cite{kurt2021lngate}. Threshold BLS can be applied in vehicular networks \cite{yang2020delegating} and communication spectrum \cite{grissa2019trustsas}. Other distributed signatures find application in various contexts, such as identity-based ring signatures for general IoT \cite{liu2023idenmultisig}, threshold EC-based signatures in wireless sensor networks \cite{sliti2008elliptic}, threshold blind signatures for data deduplication \cite{mi2020secure}, and battlefield intelligent use cases \cite{gong2021threshold}. Moreover, distributed signature verification can be used in embedded medical devices \cite{ozmen2019energy}, and distributed public key computation can transform a one-time signature to many-time for better performance \cite{behnia2021towards}. 

\underline{Electronic governance:}
Ring signatures play a significant role in e-voting \cite{li2021event,chow2008robust}, e-cash \cite{malina2018lightweight,tsang2005short}, and e-lottery \cite{kushilevitz2001fair,goldschlag1998publicly}. Additionally, both the group and ring signatures can be employed in e-commerce \cite{huang2021ba2p,camenisch2004group}.

\underline{Cloud:}
Specific schemes involving ID-based ring signatures are applied for anonymous data sharing \cite{huang2014cost} and auditable computation \cite{guo2021online}. To ensure edge computing privacy, threshold ring signatures can be utilized \cite{wang2020flexible}. Additionally, encrypted threshold digital signatures are employed for obfuscation \cite{li2021obfuscating}.

\underline{Healthcare:}
The facilitation of dynamic autonomous data interaction can be achieved by employing threshold group signatures \cite{qiao2020dynamic}. Group authorization can be accomplished through the use of threshold group Diffie-Hellman signatures \cite{nagar2017group}. Additionally, Chinese remainder-based threshold signatures can be used in medical cyber-physical systems \cite{zhou2021threshold}. 

\underline{Network:}
Network security can also benefit from applying threshold ECDSA schemes, as demonstrated by Dalskov et al. \cite{dalskov2020securing} that enhanced the security of DNS operations (DNSSEC). Their scheme protected the confidentiality of signing keys for domains, ensuring that they are not disclosed to DNS servers while safeguarding the DNS server's authenticity.

\paragraph{\textbf{Potential future directions:}}
Past efforts to construct threshold constructions have focused on two different approaches: one emphasizing a fully private approach to keep the identities of signers secret and another emphasizing accountability to facilitate the identification of signers. Certain applications, like Blockchain, demand a balance between full privacy and accountability. This necessity led to the development of a novel threshold signature scheme called Threshold, Accountable, and Private Signatures (TAPS) \cite{boneh2022threshold}. The limited existing research on TAPS presents a substantial opportunity to create implementations based on standard assumptions. Moreover, these implementations could offer features such as shorter public keys and signatures, complete traceability, and efficient verification procedures.

Unlike conventional EC-based signatures, the Schnorr signature presents numerous benefits in thresholding, such as enhanced security, efficiency, resistance to malleability, support for multi-signature functionality, etc. These advantages have been reflected in practice by the shift from ECDSA to Schnorr in Bitcoin \cite{nakamoto2008peer} and other similar systems. Considering this trajectory and the prominent works of threshold Schnorr signatures (e.g., FROST \cite{komlo2021frost}) and Schnorr multi-signatures (e.g., Musig2 \cite{nick2021musig2}), it is recommended to enhance and develop distributed Schnorr signatures with additional features (e.g., unlinkability and adaptive security) for the upcoming next-generation systems.

The emergence of federated clouds has recently affected the growth of distributed electronic healthcare systems and the expansion of IoT networks. Some of such systems employ Attribute-Based Signatures (ABSs) based on the pairings. However, their inefficiency impedes the effectiveness of data sharing in such contexts. These performance challenges in the IoT-Cloud continuum can be mitigated using the heterogeneous signatures \cite{yavuz2013eta} by offloading resource-intensive computations to capable devices.

Ongoing efforts in developing threshold PQ-secure digital signatures, despite existing inefficiencies, hold promise in creating efficient methods and reusable components. These advancements aim to facilitate the implementation of future PQ-secure signatures with structures similar to existing ones. As stated in Section \ref{sec:PQC}, NIST's fourth round signature schemes exhibit comparable structures. This suggests that the analysis methods outlined in Section \ref{sec:PQC} can be applied to these new schemes if they become standardized. We note that the ongoing standardization process is in progress, with certain schemes becoming broken while this survey is being written. Hence, it is crucial to keep an open eye on the progress of the competition while bearing in mind the adaptability of threshold methods.

%% file: bodies/conclusion.tex
This paper presents a comprehensive and systematic overview of threshold and exotic digital signatures, highlighting their pivotal role in ensuring security within the evolving landscape of decentralized next-generation networked systems and applications. We first begin by examining existing literature and identifying limitations and gaps that need to be addressed. Then, our analysis extends to both conventional and post-quantum cryptography scenarios, including custom-designed and standard signatures like NIST and NIST-PQC. In this analysis, for both conventional and PQ-secure signatures, we consider the generic methods, such as secure multi-party computation, as well as specific thresholding techniques applied across diverse signature families. Finally, we examine the application of these signatures in real-life contexts and provide a forward-looking perspective for future research.